\documentclass[conference]{IEEEtran}
\IEEEoverridecommandlockouts
\usepackage{cite}
\usepackage{amsmath,amssymb,amsfonts}
\usepackage{algorithmic}
\usepackage{graphicx}
\usepackage{textcomp}
\usepackage{xcolor}
\usepackage{eso-pic}

\usepackage{subcaption}
\usepackage{booktabs} % 如果需要更好的表格排版，通常也会用到
\usepackage{enumitem}
\usepackage{multirow} % 用于合并多行

\usepackage{soul}

\usepackage{amssymb}  % 用于 \checkmark
\usepackage{makecell} % 用于表头单元格内换行
\usepackage{pifont}
\usepackage{url}

\newcommand{\secref}[1]{Sec.~\ref{#1}}

\def\BibTeX{{\rm B\kern-.05em{\sc i\kern-.025em b}\kern-.08em
    T\kern-.1667em\lower.7ex\hbox{E}\kern-.125emX}}
\begin{document}

\AddToShipoutPictureFG*{

% Top-left acceptance text
\put(110,785){
\makebox[0pt][l]{%
\small\itshape
Accepted at the 59th IEEE/ACM International Symposium on Microarchitecture (MICRO 2026)
}
}

\put(460,750){
\includegraphics[height=1.5cm]{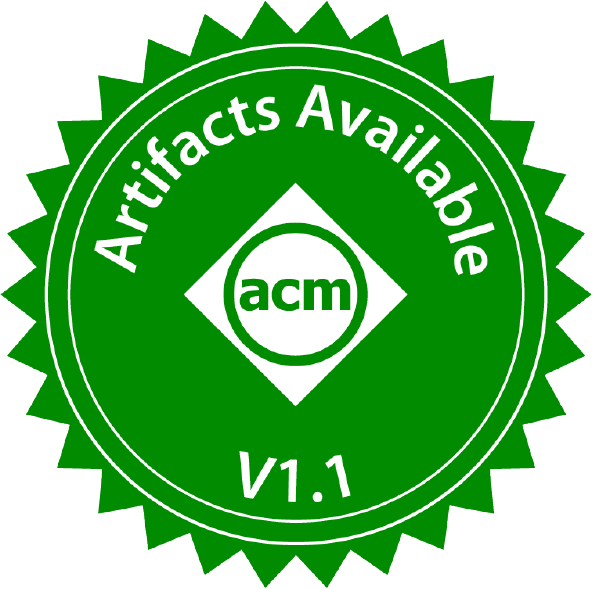}
\hspace{0.05cm}
\includegraphics[height=1.5cm]{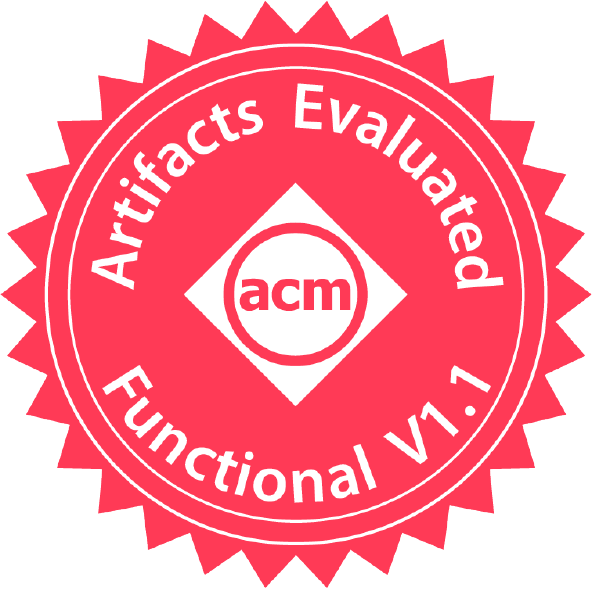}
\hspace{0.05cm}
\includegraphics[height=1.5cm]{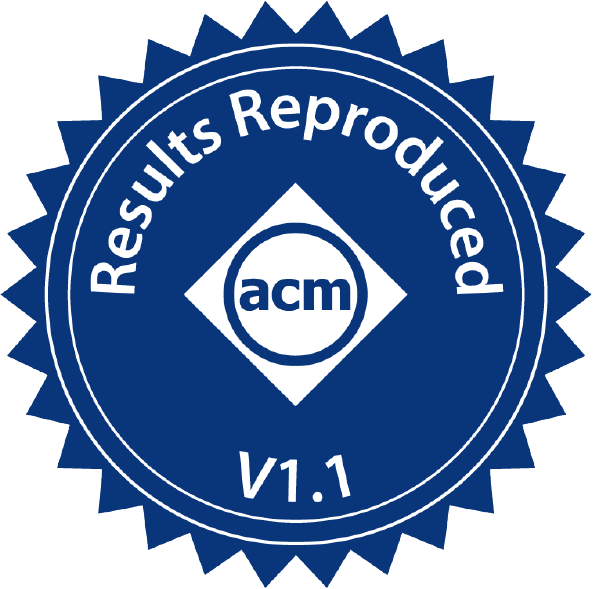}
}
}

\title{\huge HeteroReason: Heterogeneous {FPGA-GPU} Acceleration for Disaggregated Speculative Reasoning}

\author{
\IEEEauthorblockN{
Zehuan Zhang$^{1}$,
Quan Deng$^{1,2}$,
Zhibo Ren$^{1}$,
Hao Mark Chen$^{1}$,
Guoyu Li$^{1}$,
Xuchun Hu$^{1}$,\\
Jose~G.~F.~Coutinho$^{1}$,
Ce Guo$^{1}$,
Wayne Luk$^{1}$,
Zhiqiang Que$^{1,3}$,
and Hongxiang Fan$^{1}$
}

\IEEEauthorblockA{
$^{1}$Imperial College London
\quad
$^{2}$Tsinghua University
\quad
$^{3}$University of Bristol
}
}

\maketitle

\newcommand{\systemname}{HeteroReason}

\begin{abstract}

Large Reasoning Models (LRMs) have achieved state-of-the-art performance in reasoning tasks by utilizing Chain-of-Thought (CoT) reasoning. 
To achieve fast execution speed, speculative reasoning techniques have been introduced in prior work, which adopt a lightweight draft model for candidate token generation followed by process reward models (PRMs) for verification and a strong target model for refinements. 
This paper identifies that the existing speculative reasoning paradigm follows a strictly forward-only reasoning trajectory, which lacks robustness and can lead to severe error propagation if early reasoning steps are suboptimal. Furthermore, executing these disparate inference schemes, including sequential drafting and parallel verification on homogeneous GPU platforms, can lead to severe resource underutilization.
% and kernel launch overheads.
To address this, we propose {\systemname}, an algorithm-hardware co-designed heterogeneous FPGA-GPU inference paradigm specifically tailored for LRM speculative reasoning. At the algorithmic level, we introduce a backtracking-enhanced workflow that enables the system to recover from low-quality states and explore alternative reasoning trajectories, significantly improving reasoning robustness. 
At the system level, the draft model is offloaded to the FPGA while deploying the PRM and target models on GPUs. A specialized workflow is optimized to achieve prefill-decode disaggregation, which exploits shadow synchronization to overlap GPU-side refinements with FPGA-side token updates to effectively hide synchronization latency. 
To mitigate inherent sequential constraints, we propose a step-ahead speculation and refinement scheduling scheme, transitioning the system from a sequential execution scheme to a parallel pipeline. 
Experimental evaluations demonstrate that the backtracking mechanism achieves an average reasoning accuracy improvement of up to $4.2\%$ across diverse reasoning benchmarks. 
Moreover, built on the heterogeneous platforms consisting of AMD U280 or V80 FPGAs, with NVIDIA GeForce RTX 3090 GPU platforms, {\systemname} achieves $1.01\times$--$1.42\times$ latency speedups and $1.25\times$--$1.57\times$ improvements in energy efficiency under different model configurations, compared to homogeneous GPU baselines.

\end{abstract}

% \begin{IEEEkeywords}
% component, formatting, style, styling, insert
% \end{IEEEkeywords}

\section{Introduction}

LLMs have emerged as a revolutionary AI paradigm, exhibiting advanced capabilities across a wide range of applications including natural language processing~\cite{grattafiori2024llama, team2024gemma, chowdhery2023palm}, computer vision~\cite{liu2024improved,ravi2024sam,lu2024deepseek} and healthcare~\cite{yang2024advancing, liu2024survey}. Recently, the emergence of reasoning-focused LLMs, Large Reasoning Models (LRMs)~\cite{xu2025towards}, such as OpenAI o1~\cite{jaech2024openai}/o3~\cite{openai2025openai} and DeepSeek-R1~\cite{guo2025deepseek}, has led to state-of-the-art performance in reasoning tasks~\cite{li2025system,cobbe2021training,codeforces2025,gao2026model, chen2026fasttts}. Although LRMs share similar architecture backbones with conventional LLMs   and rely on autoregressive token-by-token prediction, their inference process differs: 
traditional LLMs typically aim for direct response generation~\cite{grattafiori2024llama}, while LRMs utilize Chain-of-Thought (CoT)~\cite{wei2022chain} reasoning to decompose a complex task into step-by-step reasoning sequences, yielding the final answer. This paradigm~\cite{guo2025deepseek,jaech2024openai} has demonstrated remarkable advancements in reasoning-intensive tasks.

Despite these impressive performance gains, the inference speed of this model family faces an inherent challenge: autoregressive decoding generates reasoning sequences sequentially, which introduces computational dependency and leads to a linear increase in latency with sequence length. The accumulated length of CoTs in LRMs further exacerbates this issue, thus impeding their practical deployment for real-time applications, where delayed responses degrade user experience. 
% such as voice assistants~\cite{team2024gemma}
% and autonomous robotics~\cite{lu2024deepseek}
% , where delayed responses degrade user experience.
To mitigate this issue, speculative reasoning techniques~\cite{pan2025specreason,liao2501reward} have emerged as promising solutions and have been successfully evaluated in reasoning domains.
Currently, it is a growing and active line of research spanning both academic and industry labs~\cite{liao2501reward,pan2025specreason,wang2025efficient,chu2025ssr,liu2026confspec}, with substantial inference speedups underscoring its value in latency-sensitive settings.
The core insight is that LRMs solve reasoning tasks by decomposing sequential steps, with each individual reasoning step hinging on semantic insights more than on exact tokens, enabling more tolerant approximations~\cite{pan2025specreason}.
Therefore, the techniques leverage a smaller lightweight draft reasoning model to autoregressively generate candidate tokens for individual reasoning steps, with a process reward model (PRM) or a more capable but slower target model to verify the speculative steps, guaranteeing the correct trajectory for the whole reasoning process. In this way, the target model confirms or adjusts candidate tokens from the draft model with far lower invocation frequency, thereby significantly boosting inference speed without compromising accuracy~\cite{liao2501reward}.

In the standard speculative reasoning paradigm, the algorithm adheres to a strictly forward-only reasoning trajectory in the generation process, exhibiting weak robustness and inefficiency.
For instance, if an early reasoning step yields low-quality results that are accepted to the reasoning chain, this may trigger error propagation and degrade overall accuracy since intermediate steps are interdependent.
On the hardware aspect,
draft models and target models exhibit disparate computational patterns~\cite{li2024specpim}. 
Specifically, a large portion (up to 80\%~\cite{pan2025specreason}) of the speculated steps are accepted. Therefore, in most cases, smaller draft models perform memory-intensive autoregressive decoding to generate logic sequences, whereas larger PRMs or target models conduct compute-intensive parallel verification. 
Furthermore, extensive CoTs produce a substantial amount of lightweight reasoning work, demanding frequent execution of small-scale operations for inference.
However, GPUs struggle to provide an energy-efficient solution for lightweight model inference, and frequent workload switching incurs significant kernel launch overhead.
Consequently, there is a critical need to improve algorithmic robustness and develop tailored hardware platforms to unlock the acceleration potential of speculative inference for LRMs.

\begin{figure*}[htbp]
    \centering
    \includegraphics[width=0.99\linewidth]{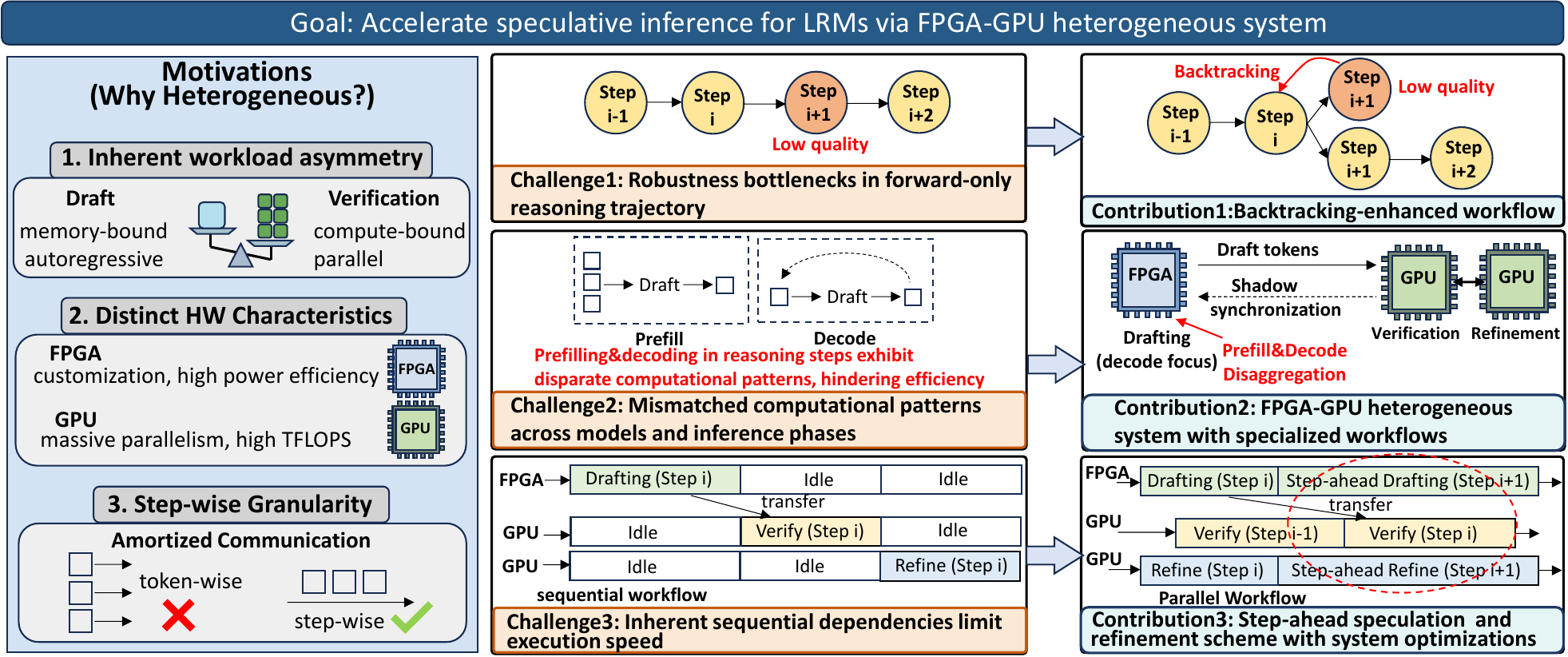}
    % \vspace{-8pt}
    \caption{ Overview of \systemname: Goals, Motivations, Challenges and Contributions.}
    \label{fig:overview_chacon}
    % \vspace{-10pt}
\end{figure*}

% GPU high power   mention edge

Although customized hardware accelerators have been investigated as shown in Table~\ref{tab:related_work},
\textit{these existing accelerators are exclusively tailored for standard token-wise speculative decoding and strictly adhere to a rigid forward-only trajectory.} They are inherently unsuited for the step-wise speculative reasoning paradigm, which entails significantly more complex computation and dynamic scheduling such as coordinating three disparate models and enabling a backtracking mechanism as detailed in~\secref{background_motivation}. 
To achieve the goal of joint algorithm-hardware optimization for LRMs, three research challenges remain:
\textbf{(1) Robustness Bottlenecks in the forward-only reasoning trajectory.} The standard speculative reasoning proceeds in a strictly forward-only manner, lacking a mechanism to recover from early-stage suboptimal generations, which severely limits reasoning robustness.
\textbf{(2) Mismatched computational patterns across models and inference phases.} The draft model prioritizes sequential autoregressive decoding, whereas the PRM performs parallel verification, and the parameter-heavy target model is invoked less frequently for refinements. 
More importantly, speculative reasoning follows a step-wise paradigm that interleaves prefilling and decoding phases. Executing these disparate patterns on a homogeneous hardware platform inevitably leads to severe inefficiency and resource underutilization.
\textbf{(3) Inherent sequential dependencies limit execution speed.
} 
The iterative process of drafting, verification, and refinements is executed in a sequential manner.
The launch of each model depends on the results from the preceding stage.
This rigid dependency restricts concurrent execution across different stages, leading to pipeline stalls that hinder further acceleration.

% \begin{table}[htbp]
% \centering
% \caption{Comparison against Existing Work.}

% \footnotesize 
% \setlength{\tabcolsep}{4pt} 
% \renewcommand{\arraystretch}{1.2} 

% \begin{tabular}{@{}c|c|c|c|c@{}}
% \toprule
% \textbf{Framework} & 
% \textbf{\makecell{Target \\ Workload}} & 
% \textbf{\makecell{Model \\ Inference Pattern}} & 
% \textbf{\makecell{BT\\ -aware}} & 
% \textbf{\makecell{Hardware }} \\ \midrule

% LP-Spec \cite{he2025lp}        & \multirow{4}{*}{\makecell{SpecDecoding \\ (token-wise)}} & Augmented Target              & \ding{55} & PIM-NPU    \\ \cmidrule{1-1} \cmidrule{3-3} \cmidrule{4-5}
% SpecPIM \cite{li2024specpim}   &                                                          & \multirow{3}{*}{Draft-Target} &    \ding{55}                           & PIM-GPU    \\ \cmidrule{1-1} \cmidrule{4-5}
% SADDLE \cite{wang2026adaptive} &                                                          &                               &       \ding{55}                        & PIM-GPU    \\ \cmidrule{1-1} \cmidrule{4-5}
% DFVG \cite{lu2026dfvg}         &                                                          &                               &     \ding{55}                          & FPGA-GPU   \\ \midrule
% Ours                  & \makecell{SpecReason \\ (step-wise)}            & Draft-PRM-Target     & \makecell{\ding{51}}   & FPGA-GPU \\ \bottomrule
% \end{tabular}\vspace{-5mm}
% \label{tab:related_work}
% \end{table}

\begin{table}[htbp]
\centering
\caption{Comparison against Existing Work.}
\label{tab:related_work}

\footnotesize
\setlength{\tabcolsep}{4pt}
\renewcommand{\arraystretch}{1.2}

\begin{tabular}{@{}c|c|c|c|c@{}}
\toprule
\textbf{Framework} &
\textbf{\makecell{Target \\ Workload}} &
\textbf{\makecell{Model \\ Inference Pattern}} &
\textbf{\makecell{BT\\ -aware}} & 
\textbf{Hardware} \\ \midrule

LP-Spec \cite{he2025lp}        & \multirow{4}{*}{\makecell{SpecDecoding \\ (token-wise)}} & Augmented Target              & \ding{55} & PIM-NPU    \\ \cmidrule{1-1} \cmidrule{3-3} \cmidrule{4-5}
SpecPIM \cite{li2024specpim}   &                                                          & \multirow{3}{*}{Draft-Target} & \ding{55} & PIM-GPU    \\ \cmidrule{1-1} \cmidrule{4-5}
SADDLE \cite{wang2026adaptive} &                                                          &                               & \ding{55} & PIM-GPU    \\ \cmidrule{1-1} \cmidrule{4-5}
DFVG \cite{lu2026dfvg}         &                                                          &                               & \ding{55} & FPGA-GPU   \\ \midrule
Ours                           & \makecell{SpecReason \\ (step-wise)}                     & Draft-PRM-Target              & \ding{51} & FPGA-GPU   \\ \bottomrule
\end{tabular}

\vspace{1mm}
\begin{minipage}{\columnwidth}
\scriptsize
BT denotes backtracking.
\end{minipage}

\vspace{-4mm}
\end{table}

To address the aforementioned challenges, we propose {\systemname}, an algorithm-hardware co-designed heterogeneous FPGA-GPU system specifically tailored to accelerate LRM speculative inference, as illustrated in Fig.~\ref{fig:overview_chacon}. 
At the algorithmic level, we introduce \textbf{(1) a backtracking-enhanced algorithm} to endow the model with backtracking capabilities, enabling it to escape suboptimal local optima and switch to alternative reasoning trajectories.
At the system level, we present \textbf{(2) an FPGA-GPU heterogeneous system with optimized workflows}. The system offloads the draft model to an FPGA and deploys the PRM and target model on GPUs, which accommodates hardware platform advantages. 
The execution workflow, featuring shadow synchronization for draft model status updates, is optimized to achieve prefill-decode disaggregation on the local FPGA accelerator.
At the scheduling level, we introduce \textbf{(3) a step-ahead speculation and refinement scheme with system optimizations.} 
To mitigate sequential constraints, this scheme allows the draft and target model to speculatively generate tokens for subsequent reasoning steps while the PRM concurrently verifies the current step, which overlaps speculation, verification and refinements, thereby minimizing pipeline stalls and reducing overall latency. 
Moreover, asynchronous target prefetching and dynamical KV cache management strategies are introduced to optimize GPU and FPGA implementations, further contributing
to system speedups.
Our main contributions can be summarized as follows: 
\begin{itemize}[leftmargin=*]
\item 
A backtracking-enhanced algorithm that enables recovery from suboptimal early generations, effectively improving the robustness and overall accuracy.~(\secref{alg_backtracking})
\item
A heterogeneous FPGA-GPU system tailored to accelerate LRM speculative inference, with the specialized workflow achieving prefill-decode disaggregation on the local accelerator via a shadow synchronization mechanism.~(\secref{hetero_system})
\item
A step-ahead speculation and refinement strategy that transitions the sequential paradigm to a parallel pipeline, integrated with system optimizations tailored for GPU and FPGA architectures, enhancing the execution speed and resource utilization.~(\secref{sec_schdule})

% \item
% Extensive Experimental Validation. We evaluate our system on an AMD V80 FPGA and NVIDIA RTX 3090 GPU platform across diverse reasoning benchmarks. Experimental results demonstrate that HeteroSpec achieves an average speedup of 1.46$\times$ (up to 1.69$\times$) and an energy efficiency improvement of 2.89$\times$ over a single GPU baseline , providing a sustainable path for high-fidelity LRM deployment.

\end{itemize}

\section{Background and Motivation}\label{background_motivation}
We introduce the background, the paradigm of reward-guided speculative reasoning (RSD) and analyze its computational characteristics and patterns. Subsequently, we elucidate the motivation for developing a heterogeneous system, and identify the challenges encountered in the efficient execution of speculative reasoning.

\subsection{Background}

\subsubsection{Large Reasoning Models (LRMs)}
LRMs simulate human-like cognition by prioritizing deliberate reasoning before response generation \cite{hu2025unveiling}.
This capability is primarily driven by the scaling of inference-time compute, which enables long CoTs for complex reasoning.
Specifically, 
LRMs break a complex problem down into intermediate steps, and then explore potential solutions and verify reasoning paths before reaching a final answer.
This paradigm demonstrates superior performance in reasoning tasks such as advanced mathematics~\cite{xu2501redstar} and medical diagnostics~\cite{huang2025o1}.
The o1~\cite{jaech2024openai} and o3~\cite{openai2025openai} model series leverage CoT reasoning, while DeepSeek’s R1~\cite{guo2025deepseek} proves the potential of reinforcement learning in developing thinking capabilities of models.
As a result, by employing explicit intermediate processing steps, structured logical reasoning capabilities are enhanced, thereby enabling effective solutions to complex problems.
However, the reasoning chains introduce substantial computational complexity, demanding higher computational costs than traditional LLMs, resulting in increased latency.

\subsubsection{Speculative Reasoning}

The speculation concept derives from computer architecture \cite{warren1985speculative}.
Speculative decoding~\cite{leviathan2023fast, miao2024specinfer, cai2024medusa, li2025eagle, chen2024sequoia, chen2024hardware} accelerates traditional LLM decoding by using a lightweight draft model to conduct faster but less accurate generation, with a strong target model verifying the candidates.
Speculative reasoning generalizes speculative decoding by enabling speculation at the step level, as the reasoning process does not require the exact reproduction of identical tokens. Unlike speculative decoding, which enforces token-level equivalence, speculative reasoning emphasizes semantic alignment. This property allows reasoning to naturally tolerate approximation~\cite{pan2025specreason}, offering substantial potential for latency reduction in LRMs.
Several explorations have been carried out.
RSD~\cite{liao2501reward} utilizes a PRM to assess intermediate steps, and the resulting rewards dynamically determine the invocation of the target model, balancing computational cost and result quality.
SpecReason~\cite{pan2025specreason} uses a smaller draft model to speculatively conduct intermediate reasoning steps and a larger base model to confirm or refine the speculated results, exploiting the semantic tolerance. 
Furthermore, the experiments demonstrate that speculative reasoning is complementary to speculative decoding, and these two optimizations can be combined to deliver further latency reductions. 
Despite these algorithmic advancements, specialized hardware support for speculative reasoning remains unexplored.

\subsection{Algorithmic  Workflow and Analysis}

\begin{figure}[htbp] % [htbp] 建议用 t (top) 保持学术排版的稳定性
  \centering
  \vspace{-2pt} 
  \begin{minipage}[t]{0.49\linewidth}
    \centering
    \includegraphics[width=\linewidth]{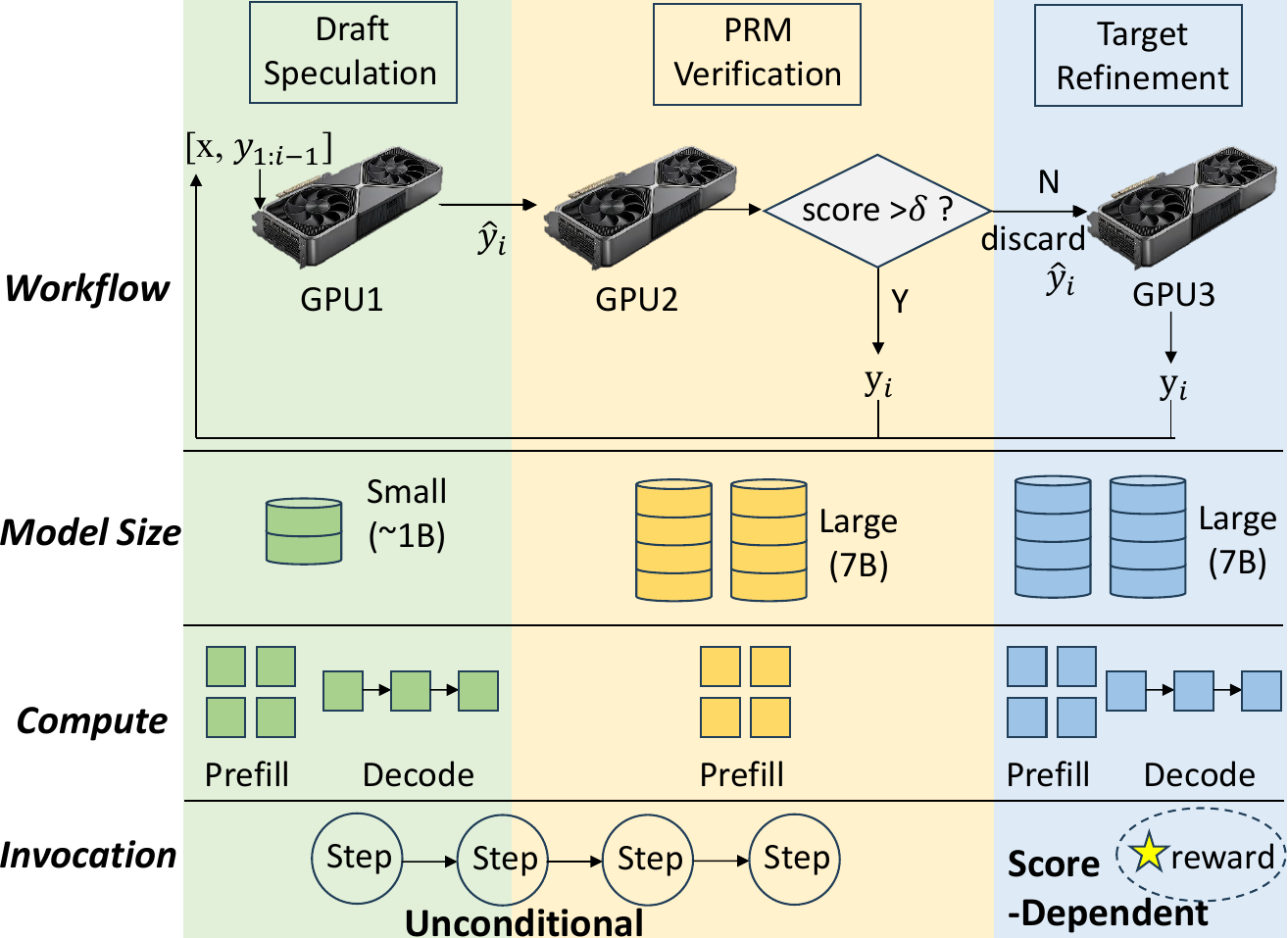}
  \end{minipage}
  \hfill
  \begin{minipage}[t]{0.49\linewidth}
    \centering
    \includegraphics[width=\linewidth]{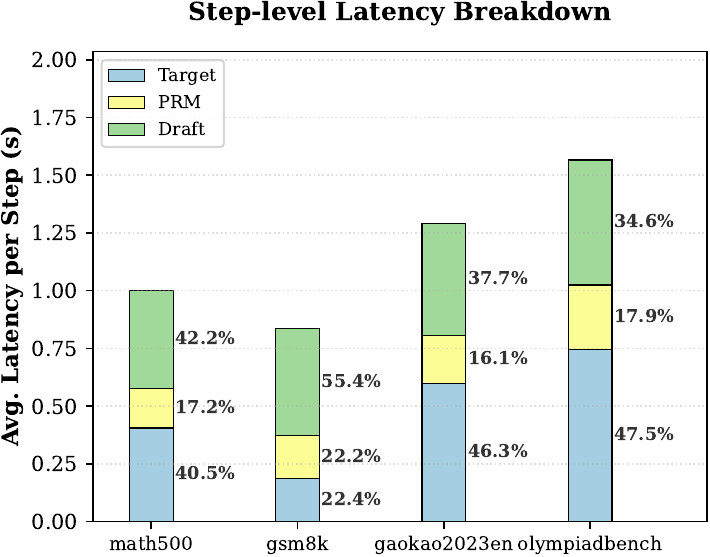}
  \end{minipage}
  \vspace{-2pt} 
  
  \caption{Overview of the speculative reasoning paradigm: the vanilla workflow (left) and the corresponding latency breakdown (right).}
  \label{fig:algorithm}
\end{figure}

The step-wise speculative reasoning paradigm operates as an iterative process as illustrated in Fig.~\ref{fig:algorithm}. 
To formalize this, let $x$ denote the initial prompt as inputs, which is processed by the draft model to generate a candidate reasoning step $\hat{y}_i$.
This candidate is then verified by the PRM. 
Subsequently,
the process branches based on the reward score.
If the reward score satisfies the criterion, $\hat{y}_i$ is accepted as the valid step $y_i$.
Conversely, if the score falls below the predefined threshold, the candidate is discarded, and the more capable target model is invoked to generate a refined step $y_i$.
Afterwards, the context is updated as $z_i = [x, y_{1:i-1}]$ serving as inputs. 
The process repeats until completion.

The algorithmic flow involves three distinct models each exhibiting varying characteristics in terms of parameter scale, compute patterns, and invocation frequency. 
As outlined in Fig.~\ref{fig:algorithm} (left), the draft model is significantly smaller (e.g., $\sim$1B parameters) than both the PRM and target models (e.g., 7B parameters). 
Regarding operational patterns, the draft model is invoked unconditionally at each reasoning step to perform speculation, necessitating both prefill and autoregressive decoding operations. Similarly, the PRM is invoked unconditionally for step verification, but it requires only prefill operations to assess the speculated sequence. Furthermore, the target model is responsible for refinements, and the invocation is score-dependent.
It is invoked only when the reward score falls below a predefined threshold. When triggered, the target model executes both prefill and decoding operations to generate a higher-quality step.

To quantify the latency breakdown, we conducted profiling experiments across four reasoning datasets to analyze the average step-level latency as shown in Fig.~\ref{fig:algorithm} (right). The results show that draft and target are the main contributors to total latency, accounting for 34.6\%--55.4\% and 22.4\%--47.5\%, respectively, while PRM contributes 16.1\%--22.2\%. 
Despite its smaller parameter size, the draft model still incurs substantial latency due to frequent unconditional invocations that require both prefill and autoregressive decoding. The target model also contributes significantly to the overall latency because of its larger model size.

\subsection{Motivation for Heterogeneous System}

Existing LLMs are predominantly deployed on high-performance GPUs, attaining remarkable performance. Step-wise speculative reasoning introduces a distinct computational paradigm that challenges the efficiency of homogeneous GPU clusters. Our proposed FPGA-GPU heterogeneous system is motivated by the following critical observations:

\subsubsection{Inherent Workload Asymmetry}

Speculative inference exhibits a fundamental decoupling between token generation and verification. The \textit{lightweight} draft model typically undergoes an autoregressive decoding process to generate candidate tokens, which is inherently sequential and \textit{memory-intensive}. 
The \textit{parameter-heavy} PRM performs parallel verification of the entire candidate sequence, which is a \textit{compute-intensive} process. 
If the verification scores fall below a threshold, the system resorts to the target model for refinements, which is \textit{parameter-heavy} but \textit{invoked  score-dependently}.
As a result, homogeneous GPU platforms struggle to efficiently accommodate the diverse computational characteristics and inherent asymmetry.
% of these models.

\subsubsection{Distinct Hardware Characteristics}

GPUs excel at parallel processing and high-throughput, arithmetic-intensive workloads, yet they come with considerable power consumption and suffer from resource underutilization and compromised efficiency when handling memory-bound tasks. 
By contrast, while FPGAs possess lower peak computational capacity than GPUs, they are characterized by superior energy efficiency and architectural reconfigurability, which enables flexible design customization tailored to specific computational demands.
The small size of the draft model makes it feasible to fit on an FPGA. Moreover, the autoregressive decoding is fundamentally bottlenecked by memory bandwidth rather than by compute capacity. By harnessing on-chip resources to develop a dedicated accelerator, the FPGA can minimize off-chip memory access overhead and achieve rapid decoding.
Consequently, the synergistic integration allows for the exploitation of the full potential of heterogeneous hardware platforms, thereby 
optimizing hardware utilization and enhancing the overall system performance.
% pushing the limits of overall system performance.

\subsubsection{Amortized Communication}

A critical bottleneck in distributed inference is the inter-device communication overhead. Completing a comprehensive reasoning task typically necessitates the generation of thousands of tokens. If communication is performed at a token-level granularity, the frequent, small-scale communication incurs prohibitive latency, severely limiting overall system throughput.
However, this bottleneck is significantly mitigated in the speculative reasoning paradigm, which performs communication at a step-wise granularity. 
In typical configurations, each reasoning step generates a sequence of fewer than 200 tokens~\cite{liao2501reward, pan2025specreason, chen2025rethinking}, enabling the effective amortization of communication overheads. For instance, in PCIe-based transfers, the communication latency of a single-step token sequence remains within the microsecond scale. Therefore, by shifting from fine-grained, token-level transfers to coarse-grained, step-level transfers, the inter-device communication costs are rendered marginal relative to the computation costs, thus facilitating efficient heterogeneous execution.
% The communication latency can be overlapped with subsequent step drafting as introduced in Section~\ref{sec:schdule}.
% Furthermore, unlike multi-process parallel execution involving resource contention, a heterogeneous configuration enables resource isolation and thus achieves conflict-free parallelism.

\subsection{Challenge Analysis}

Achieving high-performance, efficient speculative inference still faces critical challenges in both algorithmic design and hardware support.
Existing workflows lack robust error-recovery mechanisms, and no dedicated hardware systems have been developed for efficient implementations.

% \subsubsection{Robustness Bottlenecks in Forward-Only Workflow}\label{challeng1}

\subsubsection{Robustness Bottlenecks in Forward-Only Reasoning Trajectory}\label{challeng1}

\begin{figure}[htbp]
    \centering
    \includegraphics[width=0.99\linewidth]{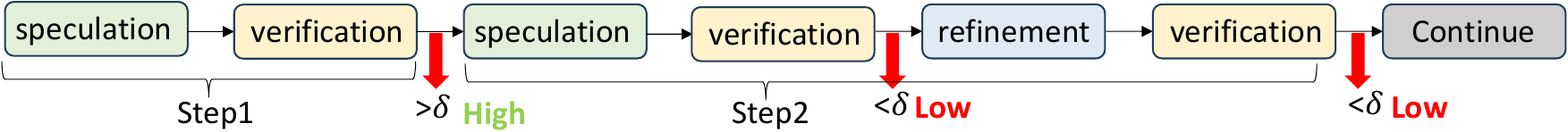}
    \caption{Robustness Bottlenecks in Forward-Only Reasoning Trajectory.}
    \label{fig:alg_forward}
\end{figure}

The vanilla workflow adheres to a strictly forward-only generation paradigm.
This "speculation-verification-conditional refinements" process introduces a critical robustness bottleneck in complex reasoning. 
If an early reasoning step receives a low
reward score but the system lacks viable alternatives, the workflow 
is forced to continue with suboptimal context.
Consider a scenario depicted in Fig.~\ref{fig:alg_forward}: The draft model generates a speculated step with a low reward score. Although the target model is invoked for refinement, it may still produce a refinement step with a low reward score due to the inherent complexity of the prompt or cumulative context noise. In such cases, the system must commit this refined yet low-quality step to the reasoning chain. Consequently, the suboptimal output becomes a permanent fixture of the context, potentially triggering error propagation across all downstream steps.

Since reasoning steps are interdependent, any committed sub-optimal steps potentially propagate errors through the entire reasoning chain.
This vulnerability is exacerbated by the inherent fidelity gap of the draft model. Due to its smaller parameter scale and the application of approximate optimization techniques (e.g., quantization), the draft model is not able to consistently guarantee high-quality outputs across diverse task difficulties.
However, the vanilla workflow lacks a mechanism to change previous potentially low-quality steps.

% \subsubsection{Aggregated Prefill-Decode}\label{challeng2}
\subsubsection{Intra-Device Prefill-Decode Interleaving of the Draft Model}\label{challeng2}

The PRM and target model are characterized by large parameter sizes and high computational intensity. The target model executes both prefill and decoding phases when triggered, while its invocation is conditional.
Consequently, both models are deployed on the GPU to leverage superior parallel processing capabilities.
For the lightweight draft model, a dedicated FPGA-based accelerator is developed for efficient speculation.
However, the current speculative reasoning paradigm necessitates a local prefill operation for the draft model at each reasoning step to continue the decoding process. 
This interleaving of prefill-decoding operations poses significant hardware design challenges due to their fundamentally distinct computational characteristics.

% This vanilla reasoning paradigm involves iterative prefill operations, which fundamentally conflicts with our design principle of decoupling compute-bound prefill from memory-bound decoding. Specifically, each reasoning step necessitates a local prefill operation for the draft model on the FPGA to continue the decoding process. 
% Forcing a dedicated decoding-optimized FPGA to handle these intermittent prefill operations introduces a severe workload-hardware mismatch. Consequently, if the FPGA performs a full-prefix prefill at each reasoning step, the cumulative latency penalty would negate the speedup gained from speculative execution.
% , transforming the low-power edge-style advantage into a performance bottleneck.

% \subsection{Limitations of Sequential Execution}
% The heterogeneous system deploys separate workloads on FPGA and GPU.
\subsubsection{Inherent Sequential Dependencies Limiting Pipeline Parallelism and Resource Utilization}\label{challeng3}
The standard speculative reasoning paradigm follows a sequential dependency: the draft model on the FPGA generates candidates, followed by the PRM and target model on the GPU for verification and refinements.
This step-wise execution leads to inter-device idle time, resulting in severe resource underutilization across the heterogeneous platform.
Specifically, the GPU remains idle while awaiting speculative tokens from the FPGA, and similarly, the FPGA stalls while awaiting feedback from the GPU.
Consequently, the total latency of each iteration equals the sum of the execution time on separate devices plus the communication overhead. To mitigate the resource underutilization and boost the speed, it is imperative to design a parallel pipeline scheduling scheme that overlaps the drafting process on the FPGA with the verification and refinements on the GPU, thereby alleviating bubbles and minimizing end-to-end system latency.

\section{Backtracking-Enhanced Reasoning}\label{alg_backtracking}

\begin{figure}[htbp]
    \centering
    \includegraphics[width=0.99\linewidth]{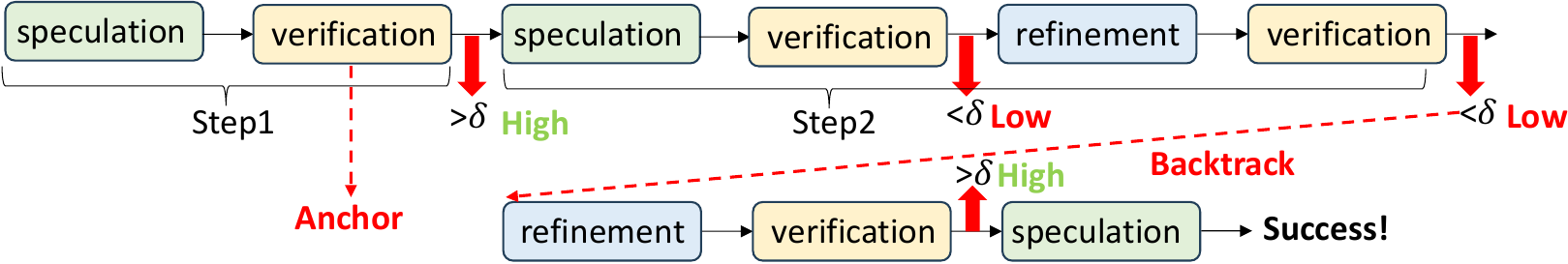}
    \caption{Backtracking Mechanism}
    \label{fig:alg_backtrack}
\end{figure}

To resolve the robustness bottleneck,
we propose a backtracking mechanism to decouple the reasoning process from suboptimal local optima. The key insight is that if both the draft and target models fail to yield high-quality results at the current step, the bottleneck likely resides in the preceding context. By maintaining a structured history of accepted states, the system can selectively backtrack to earlier "anchor points" to explore alternative reasoning trajectories. The backtracking works as follows as shown in Fig.~\ref{fig:alg_backtrack}.

The backtracking mechanism is activated only when both the draft and target models produce a low-reward step.
This ensures that backtracking is reserved for scenarios where the current context is identified as fundamentally problematic, thereby preventing unnecessary computational overhead.
Upon triggering, the system identifies valid anchors by traversing the reasoning history backward. A step is considered a valid anchor if it was previously accepted based on the draft model's speculation without target intervention. This criterion ensures that we backtrack to positions where the PRM deemed the draft output inherently satisfactory.
For a failure at step $t$, the algorithm searches for such ancestors within a lookback window of $K$ steps (e.g., $t-2, t-3$). Selecting an anchor at $t-n$ allows the system to regenerate the sequence from $t-n+1$, effectively purging the corrupted context and providing a fresh logical foundation for downstream steps.
By restricting anchors to non-intervened steps,
the workflow prioritizes the exploration of genuinely novel reasoning paths originating from draft-accepted states with high reward scores, rather than repeatedly refining the same failed trajectory.
To bound the inference latency and computational costs, the total number of backtracking attempts per problem is capped at 5, providing a principled trade-off between exhaustive search space exploration and inference efficiency.

\section{FPGA-GPU Heterogeneous System}\label{hetero_system}

% \vspace{-5pt}
\subsection{System Overview}

\begin{figure}[htbp]
    \centering
    \includegraphics[width=0.99\linewidth]{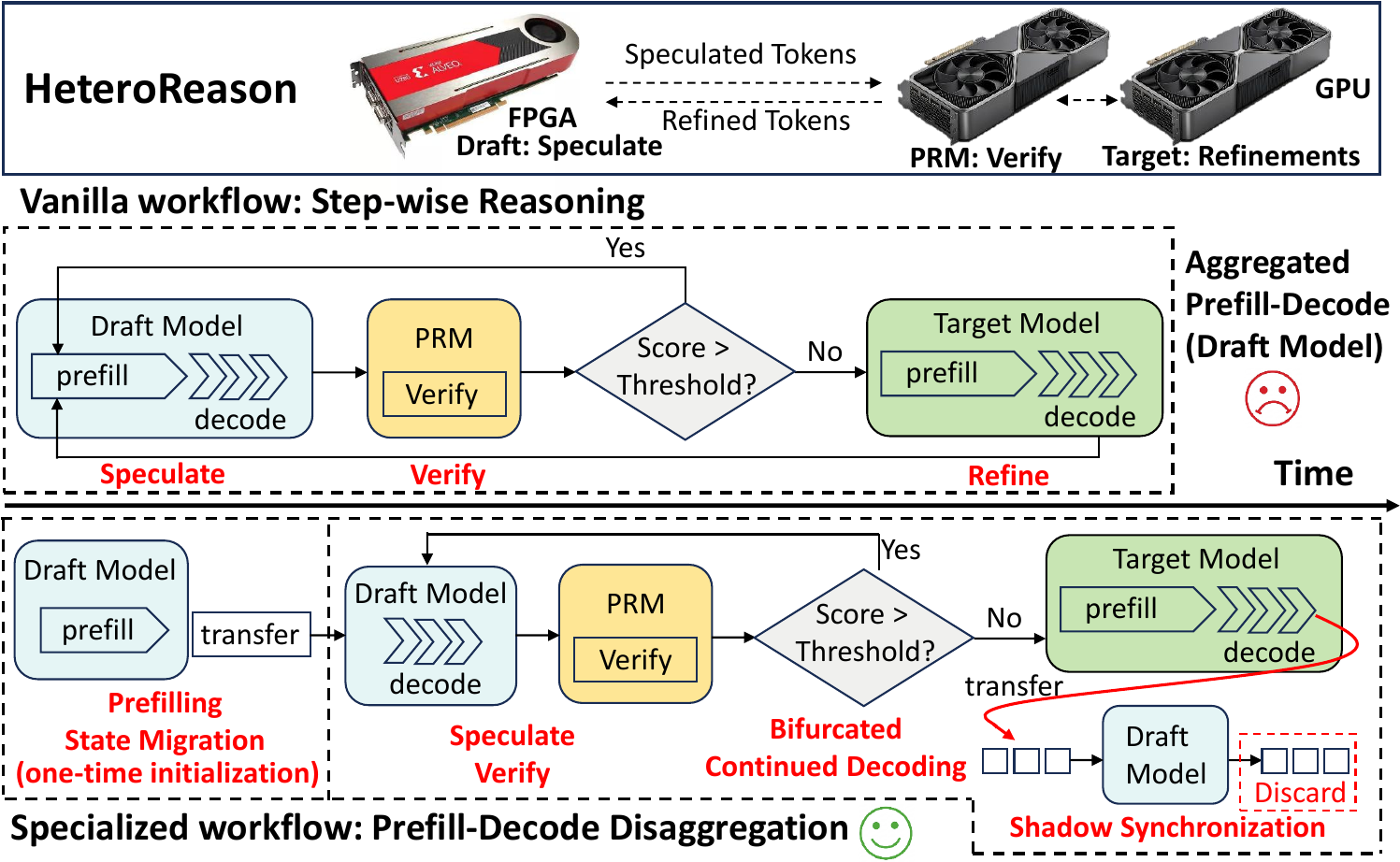}
    % \vspace{-15pt}
    \caption{System Overview of {\systemname}.}
    \label{fig:overview}
\end{figure}

The design overview of {\systemname} is illustrated in Fig.~\ref{fig:overview}, where an FPGA-based accelerator and GPUs are interconnected via the PCIe bus.
In this heterogeneous setup, the draft model is deployed on the FPGA, which is specifically optimized to minimize per-token latency during the decoding phase. 
The PRM and the target model are deployed on separate GPUs for parallel verification and refinements of the speculative sequences. These platforms exchange initial states, speculative sequences, and feedback through the PCIe.

% \vspace{-10pt}

This system achieves a synergistic workload-hardware match.
The drafting and verification workloads are disaggregated and offloaded to distinct platforms suited for the arithmetic density.
This disaggregation paradigm effectively resolves the resource underutilization caused by mismatched compute characteristics, typically encountered in homogeneous setups.
Furthermore, by exploiting a customized FPGA design for the decoding phase of the draft model, the system harnesses the advantages of heterogeneous acceleration, achieving superior energy efficiency compared to GPU-only solutions.

\subsection{Prefill-Decode Disaggregation Paradigm}
To overcome the challenges stated in Section~\ref{challeng2} and achieve disaggregation of computational patterns, a specialized workflow is optimized for the heterogeneous speculative inference system.
As illustrated in Fig.~\ref{fig:overview}, the workflow consists of three distinct phases: (1) Prefilling and State Migration, (2) Speculating and Verification, and (3) Bifurcated Continued Decoding.

\textbf{Phase 1: Prefilling and State Migration.} 
% The initial prefilling is launched  on the GPU side. Upon processing the initial prompt, the system generates the first predicted token and the corresponding Key-Value (KV) cache. Subsequently, the KV cache of the draft model is migrated from the Host memory to the FPGA's HBM via the PCIe interconnect to initialize the speculative drafting environment.
% Once initialized, the FPGA starts the iterative drafting process, executing the core layers of the draft model, leveraging its customized dataflow pipeline to minimize sequential latency.
% Due to the substantial memory and computational overhead of the large-scale vocabulary (151936 items), the intermediate hidden states from the final Transformer layer are transmitted back to the Host via PCIe.
% The Host performs the final linear projection (Logits generation) and sampling. The resulting token ID is then fed back to the FPGA to trigger the next drafting iteration.
The workflow begins with the prefilling stage on the GPU. Upon processing the initial prompt, the GPU generates the first predicted token along with its corresponding KV cache. Subsequently, the system performs state migration. The initial KV-cache for the draft model is transferred from the host memory to the FPGA's HBM via the PCIe bus. This phase initializes the speculative environment, enabling the FPGA to begin the iterative decoding process using its customized dataflow pipeline for low-latency drafting.
Note that this initialization occurs only once per reasoning task; the one-time transfer latency is rendered negligible relative to the overall generation time.

\textbf{Phase 2: Speculating and Verification.}
The FPGA executes the drafting process until the current step is completed. The resultant sequence of speculative tokens is submitted to the PRM on the GPU to conduct a single-step parallel verification.

\textbf{Phase 3: Bifurcated Continued Decoding.}
Depending on the PRM reward scores, the system dynamically branches into two execution paths to enable continued decoding.
In the case where the draft tokens satisfy the acceptance criteria,
the FPGA’s KV-cache already preserves the hidden states of both the context $z_i$ and the newly accepted tokens $\hat{y}_i$. Therefore, the draft model immediately resumes the decoding process for the subsequent step $y_{i+1}$.
This seamless continued decoding bypasses redundant prefill operations for the validated prefix.

If the draft tokens are rejected, the system enters the refinement path. The erroneous tokens are discarded, and the target model on the GPU is invoked for refinements. 
In this scenario, 
the draft model is precluded from continuing because the autoregressive decoding is strictly dependent on the integrity of the previous KV cache, but the current KV cache contains invalid states corresponding to the rejected tokens.
The target model performs a prefill operation using the validated context $z_i$ to generate refined tokens.
In the vanilla workflow, 
the FPGA-based draft model performs a local prefill to update its KV cache for the subsequent computation.

To mitigate this bottleneck, we implement shadow synchronization, which overlaps the GPU's refinement phase with the FPGA's state updates. Upon receiving a rejection signal, the FPGA immediately triggers a \textit{pointer rollback} in its KV-cache management unit, reverting the state to the last validated token. As the target model autoregressively produces refined tokens on the GPU, these tokens are streamed back to the FPGA via PCIe. Simultaneously, the draft model performs a forward pass to ingest the refined tokens and update its KV cache, discarding the output logits. Since the inference latency of the draft model is significantly lower than that of the target model, the FPGA’s state update can be fully overlapped with the refinement process on the GPU, effectively hiding the synchronization latency. This process acts as a \textit{shadow} to the refinement phase on the GPU, hence the name \textit{shadow synchronization}. This ensures that the FPGA’s drafting environment is fully synchronized and ready for the next drafting cycle by the time the target model completes its reasoning step.
Note that if the backtracking mechanism is triggered, the reasoning process reverts to the preceding step, where the target model is invoked to provide a high-fidelity alternative. In such scenarios, shadow synchronization applies as well. 
Specifically, the reverted sequence position and its corresponding KV cache pointers are transferred to the FPGA, enabling updates of internal states based on the feedback from the target model, ensuring that subsequent speculation aligns with the refined context. 
While this backtracking mechanism enhances algorithmic robustness, it inherently introduces computational redundancy. Consequently, the frequency of backtracking attempts is strictly constrained, as introduced in Section~\ref{alg_backtracking}, to balance reasoning quality with system throughput.

Overall, the specialized workflow adheres to the principle of prefill-decode disaggregation, which dedicates the FPGA-based accelerator to decoding in the drafting phase while the GPU handles parallel prefilling and verification, thus circumventing the architectural mismatch. This disaggregated approach allows for independent scaling of drafting, verification and refinements.
% resources, enabling the system to accommodate larger target models or longer sequence lengths.

% 3.3 Benefits of Prefill-Decode Disaggregation
% Our architecture strictly adheres to the principle of Prefill-Decode Disaggregation, where the computation-heavy prefilling phase is handled by the GPU/Host, while the latency-sensitive decoding (drafting) resides on the FPGA. This separation offers several advantages:

% Memory Wall Mitigation: By keeping the FPGA dedicated to decoding, we can optimize the on-chip memory hierarchy (HBM/URAM) specifically for small-batch, high-frequency accesses, avoiding the resource contention caused by high-throughput prefilling workloads.

% Architectural Specialization: The FPGA’s deterministic logic can be fully utilized for the serial dependencies of autoregressive decoding, while the GPU’s SIMT (Single Instruction, Multiple Threads) architecture is reserved for the parallel computation of prefilling and multi-token verification.

% Enhanced Scalability: This disaggregated approach allows for independent scaling of drafting and verification resources, enabling the system to handle increasing model sizes or sequence lengths without linear increases in latency.

\begin{figure*}[htbp]
    \centering
    \includegraphics[width=0.99\linewidth]{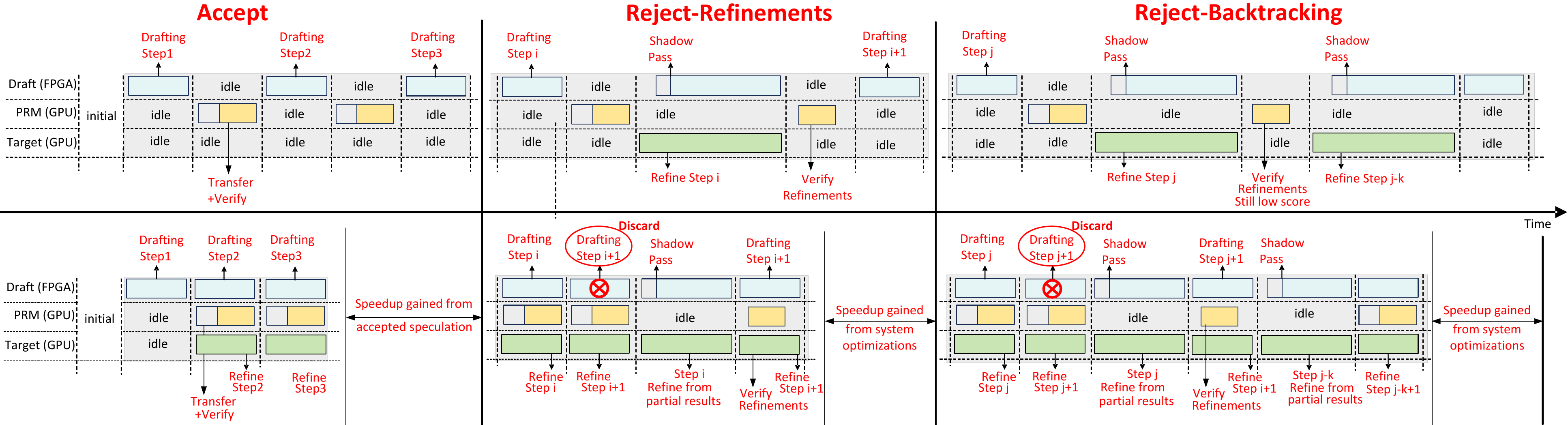}
    \caption{Step-Ahead Speculation and Refinement Scheduling Scheme.}
    \label{fig:scheduling}
\end{figure*}

\section{Step-Ahead Speculation and Refinement}\label{sec_schdule}

% \subsection{Observations and Approach}
% \subsection{Insights}

% Furthermore, recent advances in distributed speculative inference~\cite{park2025specedge} overlaps draft generation with server verification to eliminate idle time. 

% The study~\cite{pan2025specreason} demonstrates that up to $80\%$ of speculated steps are accepted, as most intermediate reasoning steps are less difficult.

To address the sequential dependency challenges identified in Section~\ref{challeng3}, we propose a step-ahead speculation and refinement strategy with system optimizations. Our core insight is that the idle states of both the draft and target models, typically caused by awaiting verification results, can be leveraged to preemptively advance the subsequent reasoning step. By initiating the next iteration concurrently with verification and inter-device communication, 
the sequential execution is transformed into a parallel pipeline, maximizing hardware utilization.
As illustrated in Fig.~\ref{fig:scheduling}, after the speculated tokens for the current step are submitted to the PRM for verification, the FPGA-based draft model and the GPU-based target model continue decoding for the subsequent step without stalling.

\subsection{Asynchronous Target Prefetching $\&$ Caching}

To boost the overall performance, an asynchronous prefetching and caching strategy is proposed for the target model on the GPU.
Specifically, the target model initiates generation for the next potential reasoning step immediately upon completion of the current step, rather than waiting for the verification outcome. By triggering the target model earlier in the pipeline, a portion of its inference latency is hidden. Since the target model is typically slower
than the draft model, it may prefetch only a partial generation. 
When prefetch execution is interrupted because a speculative step
is accepted, the generated tokens from the partial target run (without KV-cache) are cached. These cached tokens facilitate efficient
refinements and support the backtracking mechanism.
When refinement or backtracking operations are triggered, the
system reuses the partial outputs whenever the current context
shares common prefixes with the cached sequences. 
Therefore, this caching
mechanism is crucial for making refinements efficient, with which
refinement operations leverage prior computations rather than
incur full target model inference cost, effectively minimizing the
computational overhead of error recovery.

\subsection{Scheduling}
With these system optimizations, 
the proposed scheduling scheme yields speedups across three scenarios.

\textbf{Accept Scenario:} 
If the PRM validates the current step with a high reward score, the step-ahead speculation proceeds until the next speculative step is completed. Simultaneously, the step-ahead refinement continues on the GPU. 
In this case, the latency of each reasoning step is determined by the maximum of the drafting time and the verification time plus communication overhead.

% In this case, the latency associated with GPU verification and PCIe communication is effectively hidden, reducing the total step latency of the reasoning step to only the FPGA’s drafting time.

\textbf{Reject-Refinements Scenario:} 
Conversely, if the PRM assigns a low score to the current step, the step-ahead speculative tokens are immediately discarded. The step-ahead refinement continues to produce high-quality replacement tokens, while the draft model on the FPGA performs shadow synchronization to update its KV-cache using the refined tokens from the target model. Given that the draft model's inference latency is significantly lower than that of the target model, this synchronization overhead is fully overlapped, ensuring the FPGA is prepared for the next iteration without additional stalls.
In this scenario, the reuse of cached partial results contributes to the acceleration of the refinement process.

\textbf{Reject-Backtracking Scenario:} 
If the refined step still fails to satisfy the verification criteria, the backtracking mechanism is activated, reverting the process to a preceding step. The target model then updates the historical states, while the FPGA synchronizes its internal KV-cache accordingly. Similar to the refinement scenario, the speedup here is also driven by the reuse of cached partial results from the target model's preemptive execution.

Overall, by replacing idle hardware cycles with step-ahead speculation and refinement, this scheduling scheme effectively mitigates hardware bubbles and improves the utilization, leading to latency reductions across all possible scenarios.

% After the refinmements, the genrated results from the target model undergoes the verification. If verfication passes, the process continues as normal. if verfication still fails, backtracking mechanism is activated. The process reverts back to previous steps. The target model update the previous states and the draft models updates internal states through shadow synchronization.
% By replacing idle states with step-ahead speculation and refinement, 
% the idles states of FPGA and GPU are mitigated.
% Furthermore, the step-ahead speculation boosts the speed for high-score cases, while step-ahead refinement boosts the speed for low-score and backtracking cases.
% However, the discard penalty in the low-score cases prevents latency reduction for that specific step, which is determined by the capability of the draft model and PRM.

% \section{Reconfigurable Draft Accelerator}\label{sec:fpga_draft}

\subsection{Design of FPGA-based Accelerator}
In the proposed heterogeneous system, the lightweight draft model is offloaded to the FPGA side to autoregressively generate speculative tokens. 
As the draft model typically has around 1B parameters, both its weights and dynamic KV cache can be fully accommodated in the FPGA's off-chip memory.
To achieve low per-token latency, the accelerator employs a fully-streaming architecture with the hybrid mapping strategies proposed in FlexLLM~\cite{zhang2026flexllm}, 
completing the decoding process without buffering intermediate activations in off-chip memory.

\begin{figure}[htbp]
    \centering
    \includegraphics[width=0.99\linewidth]{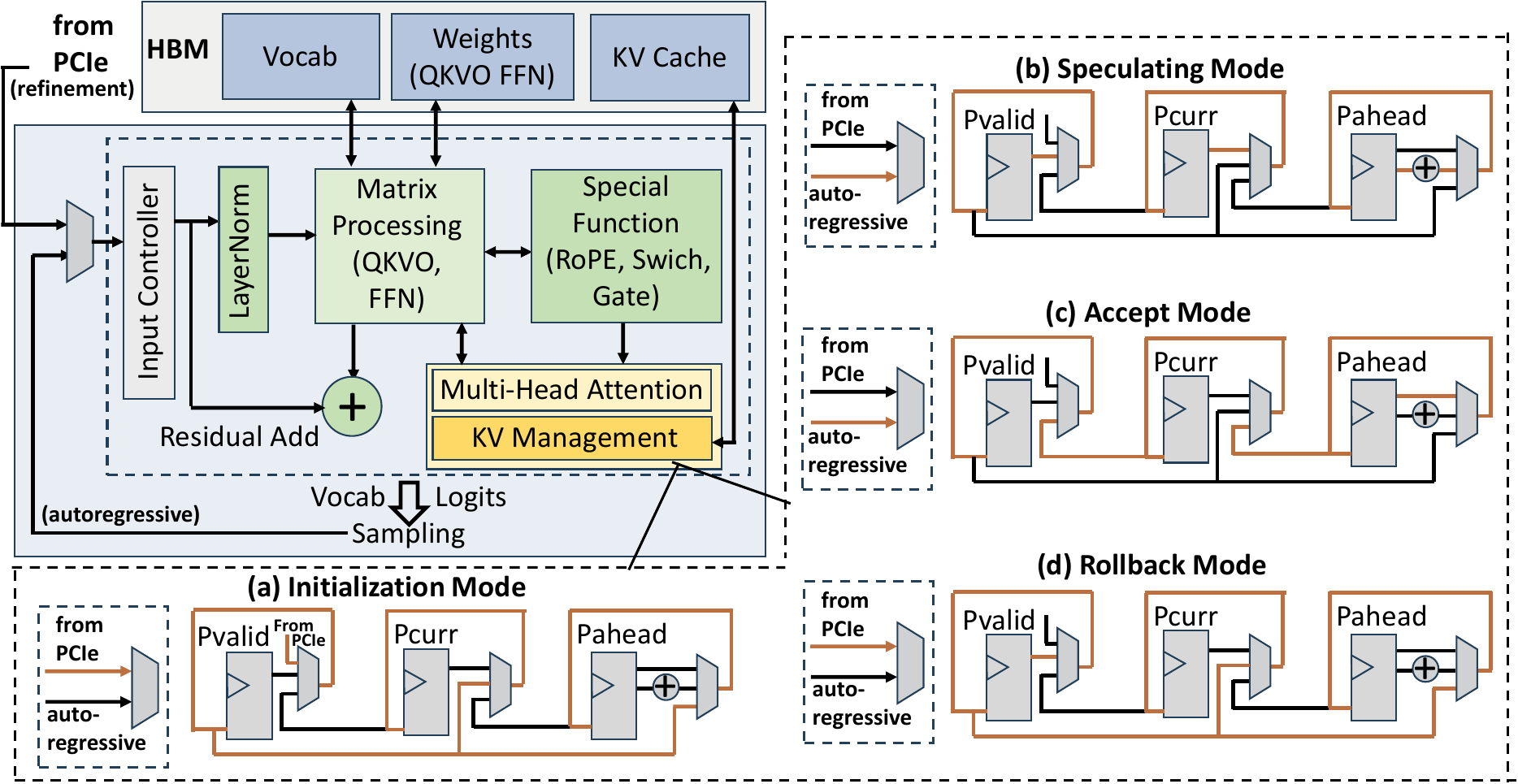}
    % \vspace{-15pt}
    \caption{Overview of the FPGA-based decoding accelerator and its four operational modes: (a) Initialization, (b) Speculation, (c) Accept, and (d) Rollback. The internal MUX and KVMU pointer logic adapt to different modes, ensuring synchronization via single-cycle state transitions.}
    
    \label{fig:FPGA_decoding}
\end{figure}

% \caption{FPGA accelerator for decoding.(a) Speculating Mode. The look-ahead pointer, $Ptr_{ahead}$, is advanced using a local incrementer. $Ptr_{valid}$ and $Ptr_{curr}$ are maintained via self-feedback. Input MUX is switched to the locally generated tokens.(b) Accept Mode.Shows the critical, single-cycle update path. $Ptr_{curr}$ is committed to $Ptr_{valid}$, while $Ptr_{ahead}$ is sourced to $Ptr_{curr}$ to advance the verified boundary.(c) Rollback Mode.Upon misprediction, the system instantly rollbacks the entire speculate history. $Ptr_{valid}$ is directly reloaded into both $Ptr_{curr}$ and $Ptr_{ahead}$ registers, effectively invalidating unverified entries. The Input MUX is redirected to source from the PCIe refined token FIFO.}

The proposed scheduling scheme requires dynamical adaptation according to reward scores from PRM.
To facilitate this responsiveness without incurring significant latency penalties, we propose a lightweight Key-Value Cache Management Unit (KVMU). As depicted in Fig.~\ref{fig:FPGA_decoding}, the KVMU employs a triplet of pointer registers to autonomously track the KV cache state: the Validated Pointer ($Ptr_{valid}$) represents the last verified context boundary, the Current Pointer ($Ptr_{curr}$) tracks the current verification step, and the Step-Ahead Pointer ($Ptr_{ahead}$) marks the step-ahead prediction frontier.

To ensure seamless KV-cache management,
the KVMU controller operates in four distinct modes, closely coupled with the scheduling phases:
\textbf{Initialization Mode}: This mode is activated during the initial system setup or upon triggering the backtracking mechanism. In this state, the three pointers ($Ptr_{valid}$, $Ptr_{curr}$, and $Ptr_{ahead}$) are reset or updated to specific addresses corresponding to the initial prompt or the reverted reasoning position. This re-synchronization ensures that the internal states are correctly restored before resuming speculative forward progress.
\textbf{Speculation Mode}: During normal autoregressive speculating, the verified boundaries ($Ptr_{valid}$ and $Ptr_{curr}$) remain stationary. Only $Ptr_{ahead}$ is advanced by the local incrementer as the draft model generates speculative tokens, pushing the frontier of the speculatively cached KV entries.
\textbf{Accept Mode}: When the FPGA receives an Accept signal, signifying a successful verification of the last speculative step, the KVMU updates pointers. $Ptr_{valid}$ is updated with the value of $Ptr_{curr}$, and $Ptr_{curr}$ is concurrently updated to the $Ptr_{ahead}$. The system then seamlessly resumes speculative forward progress.
\textbf{Rollback Mode}: Upon receiving a Reject signal, signifying a misprediction, the system instantly executes a pointer rollback to discard the invalid speculatively generated history. $Ptr_{valid}$ is directly reloaded into both $Ptr_{curr}$ and $Ptr_{ahead}$ registers.

In addition, as shown in Fig.~\ref{fig:FPGA_decoding}, the input source selection is also dynamically managed to align with the operational modes of the KVMU controller.
During \textbf{Speculation} and \textbf{Accept} modes, the accelerator maintains the autoregressive pipeline by sourcing tokens from the local sampling unit. This internal feedback loop ensures that the FPGA can continuously generate speculative tokens without the latency overhead of host-device communication. 
In contrast, \textbf{Initialization} and \textbf{Rollback} modes necessitate synchronization with the host. In these states, the input MUX is redirected to the PCIe interface, fetching high-fidelity tokens generated by the target model. 
This allows the accelerator to rapidly transition from a mispredicted speculative path to a refinement path. Thus, the subsequent drafting phase is initialized with the correct context, ensuring that the speculation process remains anchored to the high-quality reasoning chain.

This proposed KVMU design enables efficient, lightweight KV cache management that directly empowers the heterogeneous scheduling. Furthermore, the KVMU is designed as a modular, plug-and-play component, facilitating its seamless integration into existing FPGA-based LLM acceleration architectures.
\section{Experiments}

\subsection{Experimental Setup}

\textbf{Implementation and Hardware.} 
{\systemname} is evaluated on FPGA and GPU platforms. 
We consider two AMD FPGA boards, U280 and V80, and adopt NVIDIA RTX 3090 GPU platforms.
Table~\ref{table:hardware_specs} presents the specifications of the hardware platforms employed in our experiments. The FPGA designs are built on TAPA~\cite{guo2023tapa} and adopt W8A8 quantization.
We perform synthesis and P$\&$R using Vitis HLS 2022.2.
The power consumption is estimated based on implementation reports.

\begin{table}[htbp]
\centering
\caption{Hardware Specifications}
\label{table:hardware_specs}
\setlength{\tabcolsep}{2pt}
\begin{tabular*}{\columnwidth}{@{\extracolsep{\fill}}lccc@{}}
\toprule
\textbf{Metric} &
\textbf{RTX 3090} &
\textbf{U280} &
\textbf{V80} \\ \midrule

Compute units &
328 Tensor Cores &
9,024 DSPs &
10,848 DSPs \\

Frequency &
1695 MHz &
200 (300\textsuperscript{*}) MHz &
300\textsuperscript{*} MHz \\

Technology &
8 nm &
16 nm &
7 nm \\

Peak compute &
35.6 FP32 TFLOPS &
8 FP32 TFLOPS &
58 FP32 TFLOPS \\

Memory &
24 GB GDDR6X &
8 GB HBM2 &
32 GB HBM2e \\

Bandwidth &
936 GB/s &
460 GB/s &
820 GB/s \\

\bottomrule
\end{tabular*}

{\raggedright\footnotesize
\textsuperscript{*}Due to the license issue of TAPA~\cite{guo2023tapa} and RapidStream~\cite{guo2022rapidstream, guo2023rapidstream}, the FPGA performance is projected to 300 MHz.\par}

\end{table}

% \begin{table}[htbp]
%   % \vspace{-15pt}
%   \centering
%   \caption{Accelerator Resource Utilization}
%   \label{tab:resource_utilization}
%   \begin{tabular}{l S[table-format=4.1] S[table-format=4.1] S[table-format=5.0] S[table-format=4.0]}
%     \toprule
%     \textbf{Component} & {\textbf{LUT (k)}} & {\textbf{FF (k)}} & {\textbf{DSP}} & {\textbf{BRAM}} \\
%     \midrule
%     Wqkv + RoPE                 & 123.3  & 35.3  & 868  & 0    \\
%     LogFlashAttention           & 378.7  & 151.0 & 32   & 4896 \\
%     Wo + Residual               & 121.9  & 35.1  & 868  & 0    \\
%     Wup + SiLU              & 283.8  & 77.5  & 1744 & 0    \\
%     Wdown + Residual               & 468.4  & 89.0  & 1736 & 0    \\
%     LMHead                      & 153.8  & 90.1  & 1177 & 3    \\
%     Others                      & 374.4  & 203.1 & 376  & 869  \\
%     \midrule
%     \textbf{Total Design}       & \textbf{1904.2} & \textbf{681.1} & \textbf{6801} & \textbf{5768} \\
%     \textbf{Percentage}                  & \textbf{74.0\%}        & \textbf{13.2\%}       & \textbf{62.7\%}      & \textbf{77.1\%}      \\
%     \midrule
%     \textbf{Available}          & \textbf{2574.2} & \textbf{5148.4} & \textbf{10848} & \textbf{7482} \\
%     \bottomrule
%   \end{tabular}
% \end{table}

\textbf{Models.} To thoroughly evaluate the performance of our proposed system in handling diverse reasoning tasks, we employ the Qwen2.5-Math-Instruct model family~\cite{yang2024qwen2}.
The selection of the PRMs and target models follows the configurations in \cite{liao2501reward}. 
Two configurations are evaluated:
% To demonstrate the generality and scalability, three different configurations are evaluated:
(1) 0.5B draft model - 7B PRM - 7B target model
(2) 0.5B draft model - 7B PRM - 1.5B target model.
% (2): 1.5B draft model - 7B PRM - 7B target model

In the baseline configurations, the draft model, the PRM and target model are distributed across separate GPUs.
The popular framework vLLM 0.9.2 is used as the inference engine.
In our design, we replace the GPU-based draft model with our FPGA implementation and incorporate the proposed optimizations. Communication between the hardware platforms is handled via the PCIe bus.
In alignment with interactive edge computing scenarios, all experiments are conducted using a batch size of 1 to evaluate real-time inference performance.

\textbf{Hyperparameter Settings.} Following the methodology in RSD~\cite{liao2501reward}, we set both the PRM acceptance threshold and the backtracking trigger threshold to $0.7$. To bound the computational overhead during error recovery, we cap the total backtracking attempts per problem at $5$.

\textbf{Datasets.}
We conduct evaluations on four representative datasets of varying difficulty to benchmark mathematical and logical reasoning capabilities:
\begin{itemize}[leftmargin=*]
\item \textbf{Math500}~\cite{hendrycks2021measuring}: A subset of the MATH dataset consisting of 500 challenging competition-level problems.
\item \textbf{GSM8K}~\cite{cobbe2021training}: A collection of grade school math word problems requiring multi-step linguistic and numerical reasoning.
\item \textbf{Gaokao2023EN}~\cite{liao2024mario}: English-translated problems from the 2023 Chinese College Entrance Examination, reflecting high-school level proficiency.
\item \textbf{OlympiadBench}~\cite{he2024olympiadbench}: A rigorous benchmark featuring International Olympiad-level problems to test extreme reasoning limits.
\end{itemize}

\textbf{Metrics.} To comprehensively evaluate the proposed framework, we employ four key metrics to assess both algorithmic and hardware performance. 
(1) \textbf{Accuracy (\%):} Measured as the percentage of correctly answered problems across the datasets. We track this metric across different configurations to quantify the reasoning improvements brought by the backtracking mechanism, and to verify that system-level optimizations (e.g., prefetching $\&$ caching) do not compromise algorithmic fidelity. 
(2) \textbf{Latency (s/problem):} We measure the average end-to-end time taken to complete the entire reasoning process for a single problem. This metric directly reflects the user-perceived responsiveness, which is critical for real-time interactive applications. 
(3) \textbf{Goodput (Tokens/s):} We calculate goodput as the total number of accepted generated tokens divided by the end-to-end execution latency, which demonstrates the pipeline efficiency under the step-ahead speculation and refinement scheme. 
(4) \textbf{Energy Efficiency (J/Token):} The energy efficiency is defined as the total system energy consumption integrating the operating power of both the FPGA and GPU over the execution time divided by the total generated tokens.
This metric assesses the average energy cost per token and demonstrates the power savings of the FPGA-GPU heterogeneous system against homogeneous GPU baselines.

\textbf{FPGA On-Board Measurement.} We evaluate the on-board performance on
the AMD U280 FPGA running at 200 MHz. These results are used to cross-validate with our simulator and performance model. 
Achieving P$\&$R at 300 MHz requires RapidStream~\cite{guo2022rapidstream, guo2023rapidstream}, which cannot be accessed
due to licensing constraints. 
Following~\cite{zhang2026flexllm}, 
the measured performance is scaled by projecting the clock frequency at 300 MHz, and the performance on AMD V80 FPGA is estimated.

\textbf{Simulator.} To evaluate the end-to-end performance of the proposed heterogeneous architecture, we developed a specialized trace-driven simulator. The hardware execution latencies are derived from rigorous profiling: the FPGA accelerator's performance model is cross-validated with the on-board performance, while the GPU inference latencies are measured from empirical implementations. The inter-device communication overhead is modeled analytically by dividing the transferred data volume by the effective PCIe bandwidth. To ensure a fair and highly accurate evaluation, the simulator strictly replays execution traces 1:1 from the 3-GPU baseline. Specifically, we log the exact reasoning trajectory for each problem, including the total number of reasoning steps, the number of generated tokens per step, and the PRM's precise accept/reject decisions, and deterministically map them to our simulated environment. Furthermore, the simulator fully integrates our proposed step-ahead speculation and refinement scheduling strategy, accurately capturing the parallel pipeline execution strategy to calculate the final system latency and goodput.

\subsection{Algorithmic Experimental Results}

Table~\ref{tab:alg} presents the accuracy evaluation across various reasoning benchmarks under two distinct model configurations, executed on a 3-GPU baseline using BF16. The integration of the backtracking mechanism (BRSD) consistently yields notable accuracy improvements of $+2.0\%$ and $+5.0\%$ over the forward-only RSD baseline across all tested datasets. 
Furthermore, 
the complete system (BRSD + optimizations), which incorporates full system-level optimizations, maintains average accuracy gains of $+1.7\%$ and $+4.2\%$, respectively.
This demonstrates that our system optimizations effectively accelerate inference with minimal accuracy degradation ($<0.8\%$ vs. backtracking alone), while preserving the practical accuracy gains provided by backtracking.

Crucially, the experimental results reveal that the accuracy gain derived from backtracking is significantly more pronounced when employing a less capable target model. As shown in Table~\ref{tab:alg}, Config 2 with a lightweight 1.5B target model achieves a $+4.2\%$  average improvement under the fully optimized system, whereas Config 1 with a powerful 7B target model yields a $+1.7\%$ gain. This disparity indicates that, when the target model has limited capacity, its refinement generations are inherently more susceptible to logical deviations and are more likely to fail in fully correcting the draft model's errors. Under the vanilla forward-only workflow, the system is forced to commit to these suboptimal refinements, which permanently corrupt the subsequent reasoning chains. By enabling the system to escape from these suboptimal states and backtrack to alternative promising paths, this mechanism effectively compensates for the target model's capacity limitations.

% --------------------------------------------------

% --------------------------------------------------

\begin{table}[htbp]
    \centering
    \caption{Accuracy results (\%) across model configurations. \(\Delta\) shows change vs. RSD baseline. Bold indicates best per column.}

    \setlength{\tabcolsep}{4pt}
    
    % 使用 \resizebox 强制将表格宽度限制为当前单栏的宽度 (\columnwidth)
    \resizebox{\columnwidth}{!}{
    % @{} 用于去除表格最左侧和最右侧的默认空白内边距，进一步节省空间
    \begin{tabular}{@{} l r r r r r r @{}}
        \toprule 
        \textbf{Method} & \textbf{Gaokao} & \textbf{GSM8K} & \textbf{MATH500} & \textbf{Olympiad} & \textbf{Avg} & \textbf{\(\Delta\) Avg} \\

        \midrule
        \multicolumn{7}{c}{Config1: 0.5B draft - 7B PRM - 7B Target} \\

        % 按照学术规范，子表头与正文数据之间保留这条细线区隔
        \midrule 
        
        RSD (baseline) & 56.9 & 87.3 & 64.8 & 26.5 & 58.8 & -- \\
        BRSD  & \textbf{58.4} & \textbf{89.5} & 67.0 & \textbf{28.6} & \textbf{60.8} & \textbf{+2.0} \\
        BRSD + optimizations & 57.9 & 89.3 & \textbf{67.2} & 27.9 & 60.5 & 1.7 \\

        \midrule
        \multicolumn{7}{c}{Config2: 0.5B draft - 7B PRM - 1.5B Target} \\

        % 按照学术规范，子表头与正文数据之间保留这条细线区隔
        \midrule 
        
        RSD (baseline) & 42.9 & 71.3 & 50.2 & 17.0 & 45.3 & -- \\
        BRSD  & \textbf{46.0} & 78.0 & \textbf{56.8} & \textbf{20.7} & \textbf{50.3} & \textbf{+5.0} \\
        BRSD + optimizations & 45.2 & \textbf{78.6} & 54.8 & 19.7 & 49.5 & +4.2 \\
        
        \bottomrule 
    \end{tabular}
    \label{tab:alg}
    }
\end{table}

\subsection{Hardware Experimental Results}\label{sec:hardware_results}

\subsubsection{End-to-end Results}
\begin{figure}[htbp]
    \centering
    \includegraphics[width=0.99\linewidth]{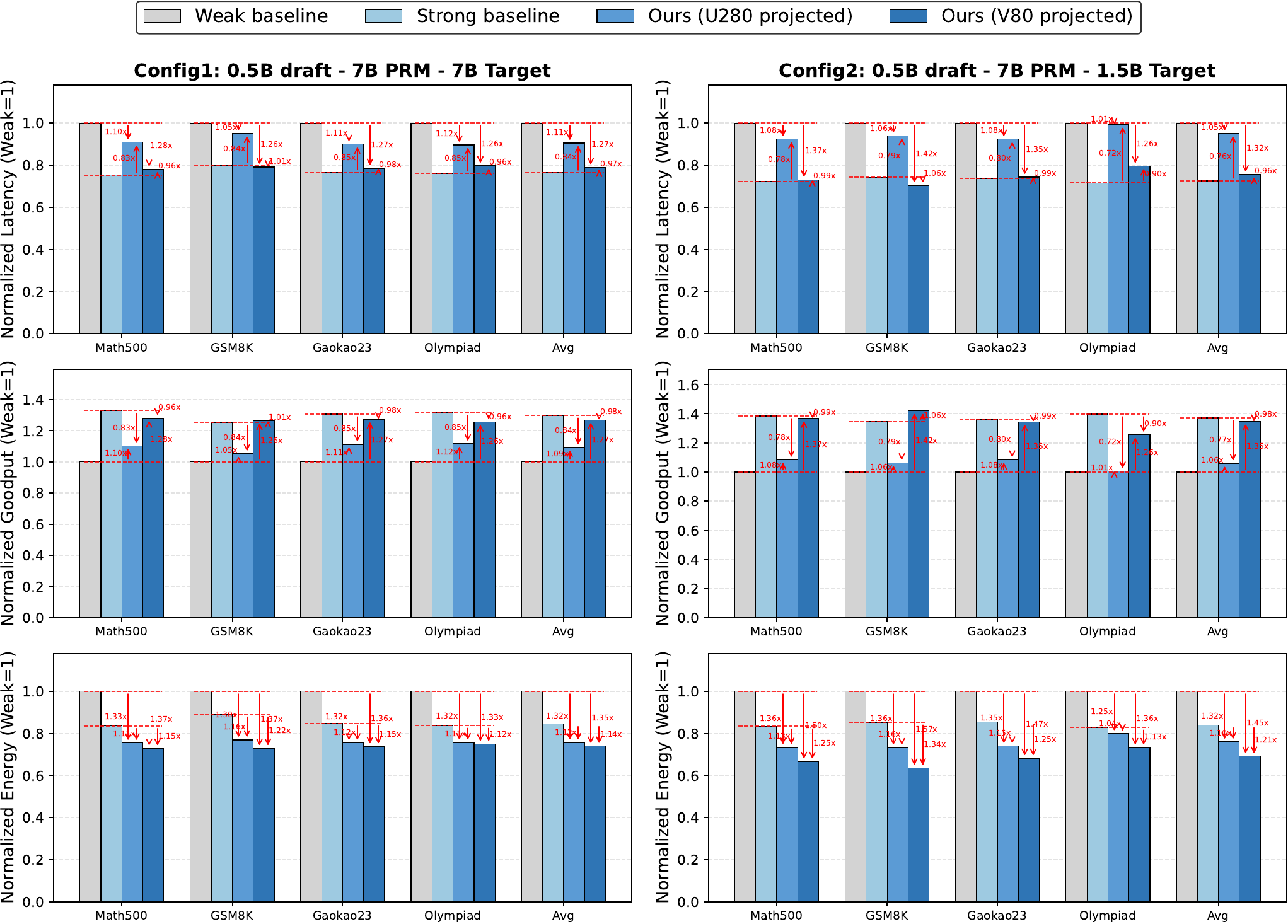}
    % {fig/5_end2end_3metric.pdf}
    % \vspace{-20pt}
    \caption{Comparisons of Latency, Goodput and Energy Efficiency against Weak and Strong Baselines.}

    \label{fig:5_end2end}
    \vspace{-10pt}
\end{figure}

Fig.~\ref{fig:5_end2end} presents a comprehensive comparison of latency, goodput, and energy efficiency between our \systemname~system and two homogeneous-GPU baselines across four diverse reasoning datasets. The weak baseline only adds backtracking to the vanilla workflow, while the strong baseline further incorporates the step-ahead speculation and refinement scheduling strategy.
Overall, our heterogeneous FPGA-GPU systems, equipped with either a U280 or V80 FPGA, both outperform the weak baseline, delivering lower end-to-end latency, higher throughput, and substantial gains in energy efficiency.
In particular, the system equipped with the V80 FPGA further narrows the performance gap with the strong GPU baseline.

% \vspace{-10pt}

\textbf{Latency Reduction.} 
As illustrated in the upper panels of Fig.~\ref{fig:5_end2end}, our heterogeneous implementations reduce latency by about $1.01\times$--$1.12\times$ (U280), and $1.26\times$--$1.42\times$ (V80) over the weak baselines, while the V80-equipped system achieves end-to-end latency comparable to the strong baseline.
This performance advantage is consistent across all benchmarks.
The speedup is primarily attributed to the prefill-decode disaggregation paradigm and the step-ahead speculation and refinement scheduling scheme, which enable a parallel pipeline associated with a decoding-only process for the draft model and reuse of generated tokens for the target model. 
In contrast, in the GPU baselines, draft models conduct both prefilling and decoding phases for each reasoning step, with the workflow bounded by strict sequential dependencies, thereby limiting execution speed.

\textbf{Goodput Enhancement.} 
The middle panels of Fig.~\ref{fig:5_end2end} exhibit the system goodput. 
Our heterogeneous implementations improve the goodput by $1.01\times$--$1.12\times$ (U280) and $1.25\times$--$1.42\times$ (V80) over the weak baselines, while the V80-equipped system achieves goodput comparable to the strong baseline.
By preemptively advancing the subsequent reasoning step, the system significantly mitigates inter-device idle states and improves hardware utilization.

\textbf{Energy Efficiency Gains.} 
The energy efficiency advantages are evaluated by comparing the consumed energy per token, as shown in the lower panels of Fig.~\ref{fig:5_end2end}. 
Our systems achieve about $1.3\times$ and $1.1\times$ higher energy efficiency than the weak and strong baselines.
These efficiency gains primarily stem from hardware-tailored optimizations on the FPGA, as its operating power consumption is significantly lower than that of the GPU.

Overall, the experimental results demonstrate that our FPGA-GPU heterogeneous architecture successfully bridges the gap between disparate computational patterns, providing a highly scalable and sustainable solution for speculative reasoning.

More experiments are conducted for further analysis.
Unless otherwise specified, the following experiments are conducted based on the heterogeneous system equipped with the U280.

\subsubsection{Comparisons with RSD}

\begin{figure}[htbp]
    \centering
    \includegraphics[width=0.99\linewidth]{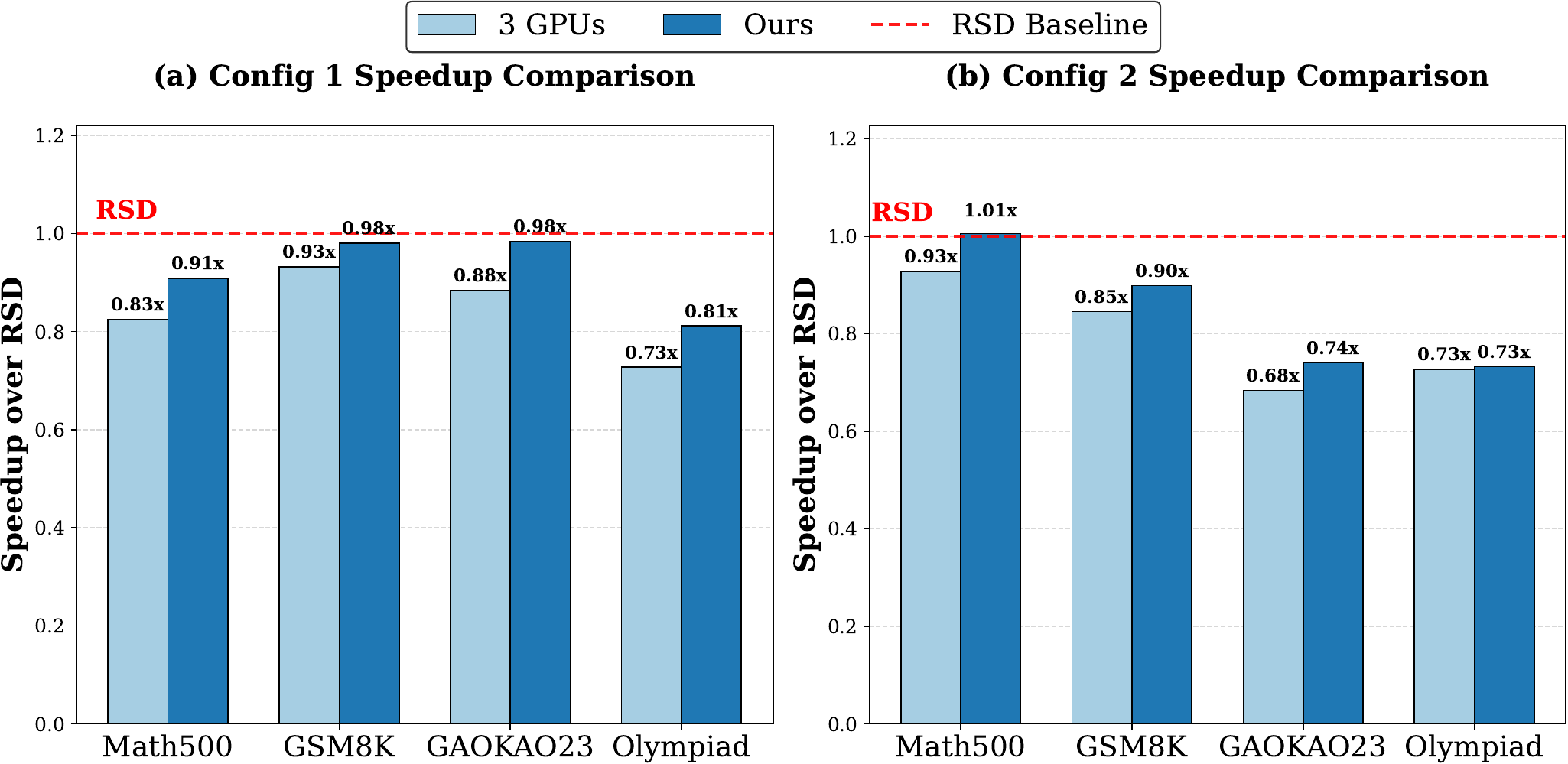}
    % {fig/5_speedup_comparison.pdf}
    \vspace{-2pt}
    \caption{ Latency Comparisons against the RSD Framework.}
    \label{fig:5_speedupRSD}
    \vspace{-10pt}
\end{figure}

To evaluate the system-level efficiency and performance overhead of our proposed backtracking-enhanced algorithm, we conduct a comparative analysis against the vanilla RSD framework. We deploy our backtracking algorithm on both a standard homogeneous 3-GPU baseline and our proposed {\systemname} system. The normalized speedups relative to the RSD baseline are illustrated in Fig.~\ref{fig:5_speedupRSD}.

When executed on a homogeneous GPU platform, the introduction of the backtracking mechanism inherently degrades execution speed. As shown in Fig.~\ref{fig:5_speedupRSD}, the 3-GPU setup consistently falls below the RSD baseline, achieving $0.73\times$ to $0.93\times$, and $0.68\times$ to $0.93\times$ the speed of RSD in Config 1 and Config 2, respectively. This performance drop highlights a critical limitation: while backtracking successfully recovers reasoning accuracy, the extra overheads impose latency penalties on standard GPU implementations.

Our proposed {\systemname} mitigates these hardware bottlenecks. Across all benchmarks and configurations, the heterogeneous system implements backtracking-enhanced approaches and accelerates the overall inference, reducing the speed degradation.

These results demonstrate the advantages of our algorithm-hardware co-design approach. By pairing the robust backtracking algorithm with a customized heterogeneous architecture, our system successfully delivers both higher accuracy and enhanced robustness while accelerating the execution speed, ultimately providing a faster and more reliable speculative reasoning system.

\subsection{Ablation Study}

\begin{figure}[htbp]
    \centering
    \includegraphics[width=0.99\linewidth]{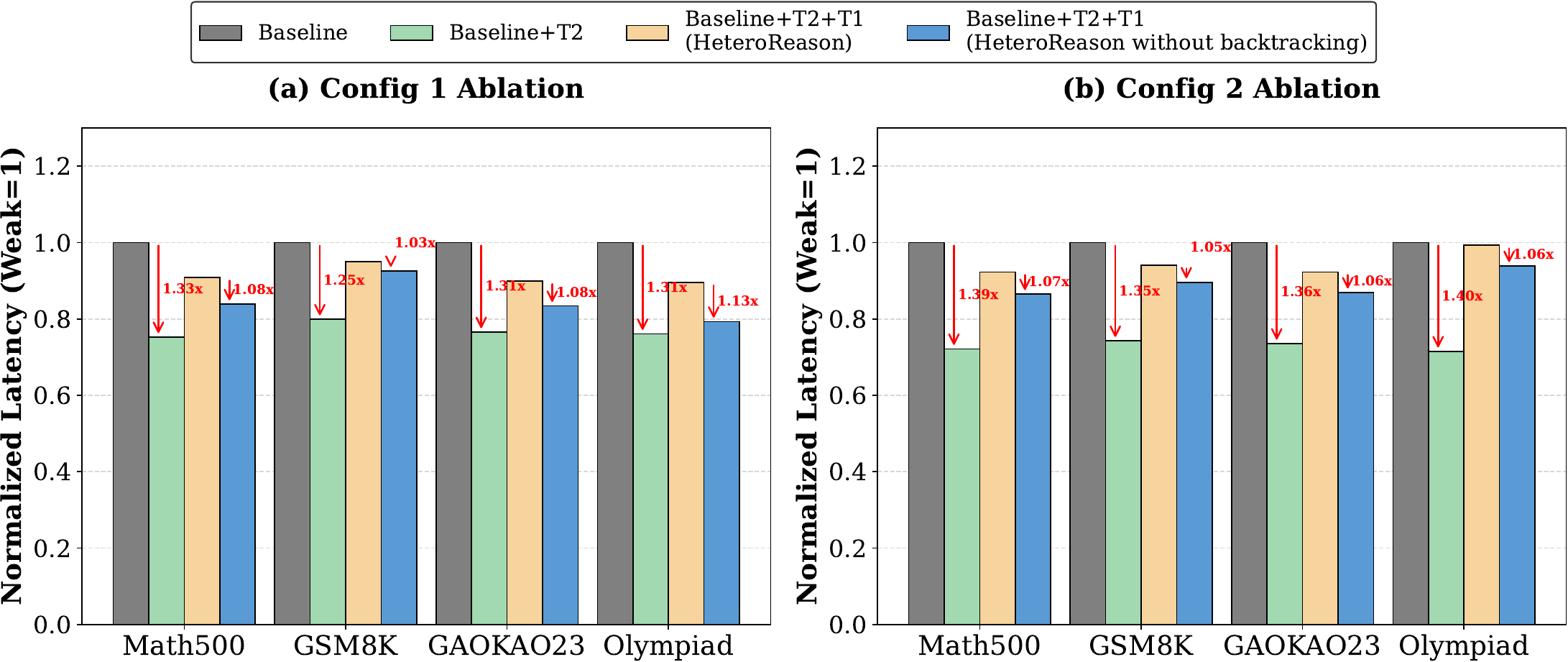}
    % {fig/5_ablation.pdf}
    % \vspace{-20pt}
    \caption{Ablation study with the step-ahead speculation and refinement scheme (T2), prefill-decode disaggregation paradigm (T1), and backtracking.}
    \label{fig:ablation}
    \vspace{-10pt}
\end{figure}

To further investigate the performance contributions of our proposed optimization techniques, we conduct an ablation study as shown in Fig.~\ref{fig:ablation}. The baseline denotes homogeneous GPU execution with backtracking enabled. (T2) adds the step-ahead speculation and refinement scheduling scheme, and (T1) further integrates the FPGA-based prefill-decode disaggregation techniques.
We also conduct experiments to analyze the impact of backtracking.

\textbf{Step-Ahead Speculation and Refinement (T2).}
The step-ahead speculation and refinement (T2) optimization implements pipeline parallelism to overlap the speculating and refinement process with the verification process.
The incorporation of the T2 scheme delivers consistent latency reductions, achieving $1.25\times$--$1.33\times$ speedup in Config 1 and $1.35\times$--$1.40\times$ speedup in Config 2 compared with the baselines. 
These results explicitly demonstrate that transitioning from a sequential paradigm to a parallel pipeline with prefetching and caching optimizations effectively mitigates inter-device idle states and boosts the overall system speed.
Note that this technique applies to diverse platforms, exhibiting high generality.

\textbf{FPGA-based Accelerator with Prefill-Decode Disaggregation (T1).} 
Building upon the T2 optimizations, the integration of T1 slightly degrades the speedup. But the heterogeneous system still shows higher energy efficiency as demonstrated in previous experiments, which is primarily attributed to lower power consumption of the FPGA.

% delivers an additional $1.09\times$--$1.83\times$ speedup in Config 1.
% This speedup advantage is primarily attributed to the fully-streaming custom acceleration architecture for the draft model and the disaggregated prefill-decode paradigm, which effectively isolates memory-bound decoding from compute-bound prefill operations. Furthermore, the lightweight dynamic KV-cache management significantly minimizes synchronization overhead, improving the overall system efficiency.

\textbf{Backtracking.}
Implementations on HeteroReason without backtracking reduce latency by $1.03\times$--$1.13\times$ in Config 1 and $1.05\times$--$1.07\times$ in Config 2. This confirms the speed-accuracy trade-off that backtracking improves accuracy but introduces additional latency overhead.
HeteroReason with backtracking enabled already outperforms GPU baselines as demonstrated in Section 6.3, and shifts this trade-off frontier.

\subsection{Communication Analysis}

Since the heterogeneous system involves communication between the FPGA and GPU, we conduct experiments to evaluate the communication costs and sensitivity.
Experimental results demonstrate that the communication costs are negligible. Across both configurations and all the datasets, the maximum communication latency accounts for only $0.0123\%$ of the end-to-end latency. These results indicate that transferring a batch of tokens in each step instead of the KV cache effectively minimizes inter-device communication overhead.

Furthermore, we quantify the speculation stall ratio, which captures cases where the FPGA completes step-ahead speculation and becomes idle while waiting for verification results from the GPU.
In the heterogeneous system, the speculation stall portion across all the steps ranges from $4.95\%$ to $26.57\%$, with an average of $17.50\%$. 
This indicates a trade-off. Faster speculation can improve overall system performance, but also increases the probability that the draft model finishes before PRM verification completes, thereby increasing the relative stall portion even if the end-to-end latency decreases.

% In summary, the occurrence of speculation stalls fundamentally depends on the interplay among the draft model's decoding speed, the inter-device communication latency, and the PRM's execution time. Our analysis demonstrates that, given a well-calibrated speculation window and appropriate bandwidth conditions, the proposed heterogeneous architecture avoids pipeline stalls and exhibits high robustness. This makes the system well-suited for diverse deployment environments, ranging from high-performance clusters to bandwidth-constrained edge-cloud collaborative systems. Furthermore, in real-world deployments, this boundary analysis provides valuable guidance for selecting appropriate model complexities and hardware platforms to ensure optimal execution efficiency.
% platforms to ensure optimal end-to-end execution efficiency.

\subsection{Energy Overheads of Wasted Tokens}

The step-ahead speculation and refinement scheme launches the next speculative execution before the current verification result is received. This aggressive scheme may introduce wasted computation when the PRM rejects the current speculative step. The prefetched tokens are then discarded because they were generated from an invalid reasoning state. To quantify this cost, we measure the additional energy consumed by these discarded tokens on the homogeneous 3-GPU platforms under Config 1 to assess the energy overheads of step-ahead scheduling.
% beyond the useful computation that contributes to the final accepted reasoning trajectory.

\begin{table}[htbp]
    % \vspace{-10pt}
    \centering
    \caption{Speedup and wasted-token energy overhead.}
    \setlength{\tabcolsep}{3pt}
    \resizebox{\dimexpr\columnwidth-2\fboxsep\relax}{!}{
    \begin{tabular}{@{} l c c c c @{}}
        \toprule
        \textbf{Dataset} 
        & \textbf{MATH500} 
        & \textbf{GSM8K} 
        & \textbf{Gaokao2023En} 
        & \textbf{OlympiadBench} \\
        \midrule
        \textbf{Speedup} 
        & $1.33\times$ 
        & $1.25\times$ 
        & $1.31\times$ 
        & $1.31\times$ \\
        
        \textbf{Overhead} 
        & $0.719\%$ 
        & $0.480\%$ 
        & $0.631\%$ 
        & $0.696\%$ \\
        \bottomrule
    \end{tabular}
    }
    % \vspace{-10pt}
    \label{tab:wasted_energy}
\end{table}

Table~\ref{tab:wasted_energy} shows the wasted-token overhead is consistently small across all evaluated datasets. The step-ahead prefetching provides a $1.25\times$--$1.33\times$ latency speedup with only $0.480\%$--$0.719\%$ additional energy consumption. {\systemname} can incur lower energy overhead due to the lower power of FPGA. These results indicate that the energy cost of discarded computation is negligible compared with the latency improvement enabled by step-ahead execution.

The low overheads are expected for two reasons. First, wasted computation is incurred only on rejected verification paths, and the rejection probability in our execution traces is typically below $30\%$. Therefore, most prefetched tokens are eventually reused by the valid reasoning trajectory rather than discarded. Second, prefetching can be interrupted. Once the reject signal is received from the PRM, the system stops the invalid step-ahead execution, which further prevents unnecessary generation after the reasoning state has already been invalidated. As a result, this scheduling scheme converts idle hardware cycles into substantial latency reduction while adding only marginal energy overhead.

\subsection{Discussion on Edge GPU/FPGA/ASIC for Speculation}

% \revhighlighttext{revShared}{The size of draft model is smaller than the PRM and the target model, making it feasible to map onto resource-constrained accelerators. Furthermore, only autoregressive decoding process is required under the proposed scheduling scheme.
% Therefore, edge GPUs, FPGAs, and ASICs are all potential platforms for executing the draft model and enabling step-ahead speculation. 
% HeteroReason deploys the draft model on the FPGA.
% Comparative analysis is conducted as follows.}

The draft model is smaller than the PRM and the target model, making it feasible to map onto resource-constrained accelerators. Edge GPUs, FPGAs, and ASICs are all potential platforms for executing the draft model and enabling step-ahead speculation and refinement. 
HeteroReason deploys the draft model on the FPGA.

\textbf{Speculation on Edge GPU.}
We evaluate the W8A8 quantized draft model on NVIDIA Jetson Orin Nano. The software stack uses Jetson-optimized PyTorch, source-built vLLM, and the FlashInfer attention backend. Compared with FPGAs, edge GPUs contain multiple parallel processing cores, but the decoding speed is about $17$~tokens/s across different prefill and decode lengths, since the decoding process is dominated by memory access rather than arithmetic throughput. 
As a result, its execution efficiency is limited by the mismatch between GPU-style SIMT parallelism and sequential token generation.
The decoding speed is even lower than the 7B target model running on the RTX 3090 GPU. Therefore, replacing the FPGA draft engine with the Jetson edge GPU would degrade the performance, which is estimated to slow down the overall system by approximately $5\times$--$6\times$ compared with the FPGA-based HeteroReason.
Consequently, moving the draft model to an edge GPU can reduce power consumption, but it does not fundamentally address the compute underutilization and memory-bound behavior of speculative drafting. By utilizing FPGA-based customized fully streaming design with dynamical KV cache management, lower latency can be achieved for these memory-intensive workloads.

\textbf{Speculation on ASIC.}
Although ASICs allow for specialized hardware optimization, they suffer from prohibitive development costs and lengthy tape-out cycles. 
In contrast, the FPGA platform offers hardware specialization while preserving reconfigurability. Adapting to new techniques typically requires generating and deploying a new bitstream rather than fabricating a new chip.

Therefore, we use the FPGA prototype to explore an optimized algorithm-hardware co-design, providing practical design insights for future ASIC implementations.

\section{Related Work}

% \subsection{Large Reasoning Models (LRMs)}

% LRMs simulate human-like cognitive processes by prioritizing deliberate reasoning prior to generating a final response, as examined in a cognitive framework~\cite{hu2025unveiling}. 
% This capability is primarily driven by the scaling of inference-time compute, which enables long CoTs for complex reasoning.

% \subsection{Speculative Reasoning}

% The speculation concept originates from the field of computer architecture~\cite{warren1985speculative}.
% Recent studies have proposed speculative decoding technique to boost the decoding speed for traditional LLMs.

% \textbf{Speculative Decoding.}
% The speculation concept derives from computer architecture \cite{warren1985speculative}.
% Recent work proposes speculative decoding to accelerate traditional LLM decoding.
% Speculative decoding works at the token-level, consisting of a speculation stage and a verification stage.
% In the speculation stage, a lightweight draft model conducts faster but less accurate generation. The generated tokens can be single sequences~\cite{leviathan2023fast} or tree structures~\cite{miao2024specinfer, cai2024medusa, li2025eagle, chen2024sequoia, chen2024hardware}. On the other hand, in the verification stage, the speculation results are verified by a larger target model concurrently. 
% In this way, latency can be significantly reduced while preserving output equivalence. 

\textbf{Algorithmic Development of Speculative Reasoning.}
Speculative decoding accelerates inference by drafting and verifying tokens strictly based on probability distributions. Recent studies on speculative reasoning have shifted towards semantic-level speculation~\cite{yang2025speculative, shi2025speccot, fu2025scaling}. 
SCoT~\cite{wang2025efficient} explores CoT-level drafting and selection and relies on task-specific LoRA alignment for the draft and selector models, unlike training-free~{\systemname}.
To mitigate rejections caused by expression divergence, where semantics are identical but tokens differ, 
the approaches in~\cite{dong2026beyond, wang2025think} probe the internal hidden states or utilize the reflective capacity to concentrate on semantic verification, enhancing the acceptance rate of speculated outputs and accelerating the execution process.
To further reduce the computational overhead introduced by the target model, a growing body of work~\cite{huang2026relayllm, fu2025r2r, maheswaran2025arbitrage, kapoor2026trim, liu2026confspec,lee2026confidence, lin2025trimr, chen2025verithinker} focuses on dynamic routing and lightweight verification. 
Recently, the speculative reasoning paradigm has demonstrated efficacy in multi-modal tasks~\cite{hu2025thinking, liu2025small} and agentic workflows~\cite{huang2026speceyes, zhong2026dualspec}. 
DREAM-R further improves multimodal speculative reasoning through RL-based refined drafting and specialized verification. Its verifier-side advances are complementary to~{\systemname}'s verifier-agnostic system mechanisms~\cite{hu2026dream}.

% SpecTemp~\cite{hu2025thinking} and SV~\cite{liu2025small} employ draft models to extract visual information for localization candidates, with a strong VLM for producing the final answer, maintaining competitive accuracy while accelerating the inference.
% SpecEyes~\cite{huang2026speceyes} and DualSpec~\cite{zhong2026dualspec} extend the speculative paradigm to action planning, employing a lightweight speculative planner to predict execution trajectories and tool-use actions, reducing the end-to-end execution time.

\textbf{Customized Accelerators for Speculative Decoding.}
Studies on hardware accelerators for speculative decoding have been conducted.
LP-Spec~\cite{he2025lp} accelerates mobile inference using GEMM-enhanced LPDDR5-PIM combined with hardware-aware token pruning.
SpecPIM~\cite{li2024specpim} utilizes architecture-dataflow co-exploration to address model heterogeneity in PIM-enabled speculative inference.
% HADES~\cite{yang2025hades} employs a modular ASIC design to specifically accelerate the verification phase.
% SPEQ~\cite{zhao2025quarter} proposes a bit-sharing quantization scheme alongside a reconfigurable processing element array to dynamically support both quantized drafting and full-precision verification.
% Mirror-SD~\cite{bhendawade2025mirror} leverages SoC heterogeneity by deploying the draft and target models on compute-dense NPUs and high-throughput GPUs, respectively. 
SADDLE~\cite{wang2026adaptive} enhances speculative decoding throughput on PIM-GPU heterogeneous systems by introducing an adaptive draft sequence length mechanism and an arithmetic intensity-aware dynamic operator scheduler.
DFVG~\cite{lu2026dfvg} proposes a heterogeneous FPGA-GPU architecture for speculative decoding, offloading the latency-sensitive draft generation to an FPGA while leveraging a GPU for high-throughput target verification.

\section{Limitations}

In~{\systemname}, we propose an algorithm–hardware co-designed heterogeneous FPGA–GPU inference paradigm for speculative reasoning. In the current implementation, some experimental results are obtained through a combination of simulation and on-board evaluation. Future work will focus on developing a fully integrated, end-to-end demonstration system. 
% Based on the existing results, the advantages of FPGA–GPU systems over GPU-only solutions are most evident under specific conditions, such as when using large target models. 
However, several current limitations may be addressed through further system-level optimization and engineering effort. From a technical perspective, the proposed methods should also be evaluated across a broader range of model architectures, including linear-attention models~\cite{shen2021efficient, dao2021transformers} and diffusion language models~\cite{lu2026adablock}, to demonstrate their generalizability.

% a lightweight KV cache management unit to facilitate rollback operations.
% Our current implementation targets transformer-based LRMs, which leverage KV caches to store historical states.
% Other model architectures, such as state space models (SSMs)~\cite{gu2023mamba, gu2021efficiently} and linear attention models~\cite{shen2021efficient, dao2021transformers}, maintain compressed latent states instead of append-only KV entries. For these models, rollback overheads are determined by their specific implementations.
% ~\cite{zhong2025specmamba} has presented efficient strategies to lower backtracking costs for Mamba.
% Accordingly, extending HeteroReason to support SSMs, linear attention models, and other model families remains feasible, which we leave as a direction for future work.

\section{Conclusion}

This work presents~{\systemname}, a heterogeneous FPGA-GPU system specifically designed to accelerate speculative inference for LRMs, offloading memory-bound, sequential drafting to FPGA while reserving GPUs for compute-intensive verification and refinements. The proposed architecture introduces three key innovations: 
a backtracking-enhanced algorithm that successfully recovers from suboptimal reasoning states to improve overall LRM accuracy, a specialized workflow with shadow synchronization to achieve prefill-decode disaggregation, and a step-ahead speculation and refinement scheduling scheme that transforms rigid sequential dependencies into a parallel pipeline. 
Experimental results across diverse reasoning benchmarks demonstrate that~{\systemname} delivers $1.01\times$--$1.42\times$ reduction in latency and $1.25\times$--$1.57\times$ gain in energy efficiency compared to homogeneous GPU baselines while improving reasoning accuracy. In conclusion, this algorithm-hardware co-design provides a highly scalable and robust solution for efficient LRM deployment.

\section*{Acknowledgment}
The support of the UK EPSRC (Grants EP/V028251/1, EP/S030069/1, and
EP/X036006/1), UKRI (Grant 256), KIAT and AMD is gratefully acknowledged.

\appendix
\section{Artifact Appendix}
\label{app:artifact_alghard}

\subsection{Abstract}
The artifact contains several components. The algorithmic
component reproduces the Table~\ref{tab:alg} accuracy results for vanilla RSD,
backtracking-enhanced RSD (BRSD), and BRSD with optimizations. The hardware
component provides a trace-driven simulator and plotting workflow for
reproducing the Fig.~\ref{fig:5_end2end} latency, goodput, and energy results from packaged
CSV/JSONL inputs without rerunning model inference.

\subsection{Artifact Check-list}
\begin{itemize}[leftmargin=*, itemsep=1pt]
  \item \textbf{Algorithm:} speculative reasoning with RSD, BRSD, and BRSD with optimized execution.
  \item \textbf{Program:} Python scripts, shell runners, step-profile CSVs, JSONL token logs, reference result summaries, and plotting utilities.
  \item \textbf{Models:} Qwen2.5 0.5B draft, Qwen2.5-Math-PRM-7B, Qwen2.5 7B target, and Qwen2.5 1.5B target. Model checkpoints are not redistributed.
  \item \textbf{Datasets:} Math500, GSM8K, Gaokao2023EN, and OlympiadBench.
  \item \textbf{Compute platform:} Algorithmic runs use 3$\times$ NVIDIA RTX 3090 GPUs in the reference setup. The packaged hardware simulator and Fig.~\ref{fig:5_end2end} plotting workflow run from CSV/JSONL traces and do not require GPUs.
  \item \textbf{Expected time:} The algorithmic smoke test takes a few minutes after model initialization; the Config1/Math500 key result takes about 1--2 hours; the full algorithmic workflow can take about 1--2 days depending on GPU availability. The packaged hardware plotting workflow takes seconds, and regenerating simulator CSVs from the packaged traces takes minutes.
  
\end{itemize}

\subsection{Description}
The artifact codebase: \url{https://zenodo.org/records/22015981} or \url{https://github.com/zehuanzhang/HeteroReason}.
The artifact is organized into several directories. \texttt{AE\_alg}
contains the algorithmic reproduction workflow.
\texttt{AE\_hardware} contains the trace-driven latency/energy simulator, the
packaged step-profile CSVs and logs, and precomputed simulator outputs.
\texttt{AE\_hardware\_V80} is used for V80 simulations.

The three algorithmic methods in Table~\ref{tab:alg} are:
\begin{itemize}[leftmargin=*, itemsep=1pt]
  \item \texttt{rsd}: the vanilla speculative reasoning baseline.
  \item \texttt{brsd}: RSD with backtracking.
  \item \texttt{brsd\_optimized}: BRSD with the optimizations.
\end{itemize}
These correspond to the implementation names \texttt{beam1},
\texttt{bbeam1}, and \texttt{bbeam1\_prefetch\_cache}, respectively.

\subsection{Installation}
After downloading the artifact, follow the installation instructions in the
\texttt{README.md}, which describes the Python dependencies,
the local model paths, and
the environment variables required for algorithmic reproduction.
The hardware simulator and plotting workflow use
packaged CSV/JSONL inputs and do not require model checkpoints or GPU
inference.

\subsection{Algorithmic Experiment Workflow}
The algorithmic reproduction workflow is documented in the
\texttt{README.md}. It provides three reproduction levels: a smoke test on
eight fixed Config1/Math500 examples, the Config1/Math500 key-result runs,
and the full experiments in Table~\ref{tab:alg}.
The smoke test is intended as the first validation step. The key-result
workflow focuses on Config1/Math500, while the full workflow reproduces the
complete algorithmic results.

\subsection{Hardware Simulator and Figure Workflow}
The \texttt{README.md} also documents hardware simulator and
plotting workflow. The results can be regenerated without rerunning model inference or the
hardware experiments.

\subsection{Evaluation and Expected Results}
The algorithmic smoke test reports per-method accuracy and latency on the fixed
eight-example subset. The key-result workflow reproduces the Config1/Math500
Table~\ref{tab:alg} entries, with \texttt{brsd\_optimized} as the recommended artifact
result. The full workflow reproduces all Table~\ref{tab:alg} algorithmic accuracy values
across Config1 and Config2.

The hardware workflow regenerates Fig.~\ref{fig:5_end2end} and the corresponding
CSV/JSON metric summaries. The
expected metrics include average latency, goodput, and average energy per
problem.

\subsection{Notes}
Algorithmic latency is platform-dependent and can vary with GPU
types. The hardware simulator
and plotting workflow are deterministic for the supplied CSV/JSONL inputs. Model
checkpoints are referenced by local paths or Hugging Face identifiers and are
not redistributed with the artifact.

\bibliographystyle{IEEEtran}
\bibliography{./text/reference}

\end{document}